%% file: ratarXiv4.tex
\documentclass[10pt]{article}

\usepackage{amsmath, amssymb, latexsym}
\input{mymacs}

\def\llangle{{\langle \kern -.2em \langle}}
\def\rrangle{{\rangle \kern -.2em \rangle}}
\def\rat{{\rm  rat}}
\def\rec{{\rm  rec}}

\def\Rat{\hbox{\bf Reg}}
\def\supp{{\rm supp}}

\def\k{{\mathbf{k}}}
\def\parallel{{|\kern -.1em |}}

\def\N{\mathbb{N}}
\def\B{\mathbb{B}}
\def\Z{\mathbb{Z}}
\def\bA{\mathbf{A}}
\def\bB{\mathbf{B}}
\def\bC{\mathbf{C}}
\def\W{{\mathcal{W}}}

\def\b1{{\mathbf{1}}}

\begin{document}

\title{\textbf{Axiomatizing Rational Power Series}}

\author{
S. L. Bloom\\
Dept. of Computer Science\\
Stevens Institute of Technology\\
Hoboken, NJ, USA\\
\and
Z. \'Esik\thanks{Partially supported by grant no. MTM2007-63422 from the 
Ministry of Education and Science of Spain.}\\
Dept. of Computer Science\\
University of Szeged\\
Szeged, Hungary \\
GRLMC\\
Rovira i Virgili University\\
Tarragona, Spain
}

\date{\empty}

\maketitle

\begin{abstract}
Iteration semi\-rings are Conway semi\-rings satisfying 
Conway's group identities. 
We show that the semi\-rings $\N^{\rat}\llangle \Sigma^* \rrangle$ 
of rational power series with coefficients in the semi\-ring $\N$ 
of natural numbers are the free partial iteration semi\-rings. Moreover,  
we characterize the semi\-rings  $\N_\infty^{\rat}\llangle \Sigma^* \rrangle$ 
as the free semi\-rings in the variety of iteration semi\-rings 
defined by  three additional simple identities, where $\N_\infty$ 
is the completion of $\N$ obtained by adding a point of infinity. 
We also show that this latter variety coincides with the variety generated by 
the complete, or continuous semirings. As a consequence of these results, we  
obtain that the semi\-rings $\N_\infty^{\rat}\llangle \Sigma^* \rrangle$,
equipped with the sum order, are free in the class of symmetric inductive 
$^*$-semi\-rings. This characterization corresponds to Kozen's axiomatization 
of regular languages.  
\end{abstract}

\section{Introduction}

One of the most basic algebraic structures studied in Computer Science
are the semi\-rings $\Rat(\Sigma^*)$ of regular (or rational) 
languages over an alphabet $\Sigma$ equipped with the star operation. 
Salomaa \cite{Salomaa} has axiomatized these semi\-rings of 
regular languages using
a few simple identities and the \emph{unique fixed point rule} 
asserting that if the regular language $a$ does not contain the empty 
word then $a^*b$ is the unique solution of the fixed point equation 
$x= ax + b$. There are several ways of expressing the empty word 
property using a first-order language. Probably, the simplest way is 
by the inequality $1 + a \neq a$. 
Using this, the unique fixed point rule can be formulated 
as the first-order axiom 
\begin{eqnarray*}
\forall a \forall b \forall x ((1 + a \neq a \wedge ax + b = x)
 \Rightarrow x = a^*b).
\end{eqnarray*}
Salomaa's result then amounts to the assertion that for any $\Sigma$,
$\Rat(\Sigma^*)$ is freely generated in the class  of $^*$-semi\-rings 
satisfying a finite number of (simple) identities and the above axiom.
We have thus a \emph{finite} first-order axiomatization of regular
languages.

Because of the extra condition on $a$, the 
unique fixed point rule is not a quasi-identity.
A finite axiomatization using only quasi-identities 
has been first obtained by Archangelsky and Gorshkov, cf. \cite{AG}.
A second, and perhaps more serious concern is that several natural 
$^*$-semi\-rings which satisfy all identities of regular
languages are not models of the 
unique fixed point rule. Examples of such semi\-rings 
are semi\-rings of binary relations with the  
reflexive-transitive closure operation as star, since for 
binary relations, the equation $x =ax +b$ usually has several solutions, 
even if $1 + a \neq a$ (i.e., when $a$ not reflexive). 
On the other hand, $a^*b$ is least among all solutions, so that
$$\forall a \forall b \forall x (x = ax + b \Rightarrow a^*b \leq x)$$
where $a^*b \leq x$ may be viewed as abbreviation for  $a^*b + x =x$. 
And indeed, the semi\-rings of regular languages can be characterized
as the free algebras in a quasi-variety of semi\-rings with a star 
operation axiomatized 
by a finite set of simple identities and the above 
\emph{least fixed point rule}, or the \emph{least pre-fixed point rule}
$$\forall a \forall b \forall x (ax + b \leq x \Rightarrow a^*b \leq x).$$
This result is due to Krob \cite{Krob}. In \cite{Kozen90,Kozen},
Kozen also required the dual of the least (pre-)fixed point rule
$$\forall a \forall b \forall x (xa + b \leq x \Rightarrow ba^* \leq x),$$
and gave a simpler proof of completeness of this system.
Several other finite axiomatizations are derivable from 
Krob's and Kozen's systems, see \cite{Boffa1,Boffa2,BEbul}.

But the largest class of algebras in which the semi\-rings of 
regular languages are free is of course a variety. This 
variety, the class of all semi\-rings with a star operation satisfying all 
identities true of regular languages, is the same as
the variety generated by all $^*$-semi\-rings of binary relations.
The question whether this variety is finitely based 
was answered by Redko \cite{Redko1,Redko2} and Conway \cite{Conway},
who showed that there is no finite (first-order or equational) 
axiomatization. The question of finding infinite equational 
bases was considered in \cite{BEreg,Krob}. The system given in 
Krob \cite{Krob} consists of the Conway semi\-ring identities, 
the identity $1^* = 1$, and Conway's 
group identities \cite{Conway} associated with the finite (simple) groups.
Conway semi\-rings were first defined formally in \cite{BEmat,BEbook}. 
Conway semirings are semi\-rings equipped with a star operation 
satisfying $(a+b)^* = a^*(ba^*)^*$ and $(ab)^* = a(ba)^*b + 1$.
Conway semi\-rings satisfying the infinite collection 
of group identities are called iteration semi\-rings, cf. \cite{Esgroup}.
The terminology  is due to the fact that iteration semi\-rings are 
exactly the semi\-rings which are iteration algebras, i.e., satisfy 
the axioms of iteration theories \cite{BEbook} which capture the 
equational properties of the fixed point operation. Thus, 
Krob's result characterizes the semi\-rings of regular languages 
as the free iteration semi\-rings satisfying $1^* = 1$ (which implies that 
sum is idempotent). Another proof of this result using iteration theories 
can be obtained by combining the axiomatization of 
regular languages from \cite{BEreg} and the  
completeness (of certain generalizations of) the group 
identities for iteration theories, established
in  \cite{Esgroup}.

In this paper, we drop the idempotence of the sum operation 
and consider the semi\-rings of rational power series 
$\N^{\rat}\llangle \Sigma^* \rrangle$ 
and $\N_\infty^{\rat}\llangle \Sigma^* \rrangle$ over the semi\-ring 
$\N$ of natural numbers and its completion $\N_\infty$ with a point 
of infinity. The star operation in $\N^{\rat}\llangle \Sigma^* \rrangle$ 
is defined only on those proper power series having $0$ as the coefficient 
of the empty word (the empty word property), whereas the star operation 
in $\N_\infty^{\rat}\llangle \Sigma^* \rrangle$ is totally defined. 
We prove that $\N_\infty^{\rat}\llangle \Sigma^* \rrangle$ is freely generated 
by $\Sigma$ in the variety $\V$ of all iteration semi\-rings satisfying the identities 
$1^* 1^* = 1^*$, $1^*a = a1^*$ and $1^*(1^*a)^* = 1^*a^*$. 
This result is also of interest because $\V$ coincides with the variety 
generated by those $^*$-semirings that arise from (countably) complete 
or continuous semirings by defining $a^*$ as the sum $\sum_{n \geq 0} a^n$. 
Moreover, we prove that $\N^{\rat}\llangle \Sigma^* \rrangle$ is freely generated by 
$\Sigma$ in the class of all \emph{partial} iteration semi\-rings. 
As a consequence of the equational axiomatizations, we 
show that $\N_\infty^{\rat}\llangle \Sigma^* \rrangle$, equipped with the 
the sum order, is freely generated by $\Sigma$ in the class of 
ordered $^*$-semi\-rings satisfying the fixed point identity 
$aa^* + 1 = a^*$ and the least pre-fixed point rule.

The paper is organized as follows. In Section~\ref{sec-semirings}  we review 
the notion of semi\-rings and power series. Section~\ref{sec-Conway} is devoted 
to (partial) Conway and iteration semirings. In Section~\ref{sec-Kleene} we provide a 
formulation of the Kleene-Sch\"utzenberger theorem for (partial) Conway semirings 
from \cite{BEbook,Espartial}. In the characterization of the semirings 
$\N^{\rat}\llangle \Sigma^* \rrangle$ as the free partial iteration semirings,
in addition to the Kleene-Sch\"utzenberger theorem, our main tool will be the 
commutative identities. We establish several technical results for the 
commutative identity in Section~\ref{sec-commutative}. Section~\ref{sec-Nrat} is 
devoted to proving the freeness result for the 
semirings $\N^{\rat}\llangle \Sigma^* \rrangle$ mentioned above.
Then, in Section~\ref{sec-Ninftyrat1} we prove that the semirings 
$\N_\infty^{\rat}\llangle \Sigma^* \rrangle$ are free in the 
variety if iteration semirings satisfying three additional simple 
identities. Last, in Section~\ref{sec-Ninftyrat2} we characterize the semirings 
$\N^\rat_\infty\llangle \Sigma^* \rrangle$ as the free symmetric inductive 
$^*$-semirings, and as the free inductive $^*$-semirings satsifying 
an additional inequation.

\section{Semi\-rings}
\label{sec-semirings}

A \emph{semi\-ring} \cite{Golan} is an algebra $S = (S,+,\cdot,0,1)$ 
such that $(S,+,0)$ is a commutative monoid, where $+$ is called sum 
or addition, $(S,\cdot,1)$ is a monoid, where $\cdot$ is called 
product or multiplication. Moreover, $0$ is an absorbing element 
with respect to multiplication and product distributes over sum:
\begin{eqnarray*}
0 \cdot a &=& 0\\
a \cdot 0 &=& 0\\
a(b+c) &=& ab + ac\\
(b+c)a &=& ba + ca 
\end{eqnarray*}
for all $a,b,c \in S$. 
A semi\-ring $S$ is called \emph{idempotent} if 
\begin{eqnarray*}
a + a &=& a
\end{eqnarray*}
for all $a\in S$. A morphism of semi\-rings preserves the sum and product 
operations and the constants $0$ and $1$. Since semi\-rings are defined 
by identities, the class of all semi\-rings is a variety (see e.g., \cite{Gratzer})
as is the class of all idempotent semi\-rings. 

An important example of a semi\-ring is the semi\-ring $\N = (\N,+,\cdot,0,1)$ 
of natural numbers equipped with the usual sum and product operations.
An important example of an idempotent semi\-ring is the boolean 
semi\-ring $\B$ whose underlying set is $\{0,1\}$ and whose sum and 
product operations are the operations $\vee$ and $\wedge$, i.e., 
disjunction and conjunction. Actually $\N$ and $\B$ are respectively the 
initial semi\-ring and the initial idempotent semi\-ring.
The semiring $\N_\infty$ is defined on the set $\N \cup\{\infty\}$
so that it contains $\N$ as a subsemiring and $n + \infty = \infty +n = \infty$
and $m \infty = \infty m = \infty$ for all $n,m \in \N \cup \{\infty\}$, $m \neq0$.

We describe two constructions on semi\-rings. For more information on semi\-rings, 
the reader is referred to Golan's book
\cite{Golan}.  

The first construction is that of matrix semi\-rings. 
When $S$ is a semi\-ring, then for each $n \geq 0$ 
the set $S^{n \times n}$ of all $n \times n$ matrices 
over $S$ is also a semi\-ring. The sum operation is 
defined pointwise and product is the usual matrix product.
The constants are the matrix $0_n$ all of whose entries 
are $0$ (often denoted just 0), and the diagonal matrix 
$E_n$ whose diagonal entries are all $1$. In addition 
to square matrices, we will also consider more general 
rectangular matrices with the usual definition of sum 
and product. (Rectangular 
matrices over $S$ form a semiadditive category that 
can canonically be assigned to $S$ but we will avoid using 
categorical notions.) When $\rho$ is a function 
$\{1,\cdots,n\} \to \{1,\cdots,p\}$, for some $n,p\geq 0$,
there is a corresponding $n \times p$ matrix over 
each semi\-ring: it is  a $0$-$1$ matrix with a $1$ on the 
$(i,j)$th position exactly when $i\rho = j$. Such 
matrices will be called \emph{functional}. A \emph{permutation 
matrix} is a functional matrix that corresponds to a permutation. 

The second construction is that of power series and polynomial 
semi\-rings, cf. \cite{Berstel}. 
Suppose that $S$ is a semi\-ring and $\Sigma$ is a set. Let $\Sigma^*$ 
denote the free monoid of all words over $\Sigma$ including the 
empty word $\epsilon$. A \emph{power series} over $S$ (in the noncommuting 
letters in $\Sigma$) is a function 
$s: \Sigma^* \to S$. It is a common practice to represent a power series 
$s$ as a formal sum $\sum_{w \in \Sigma^*}(s,w)w$, where the 
\emph{coefficient} $(s,w)$ is $ws$, the value of $s$ on the word $w$.
The \emph{support} of a series $s$ is the set $\supp(s) =
\{w : (s,w)\neq 0\}$. When $\supp(s)$ is finite, $s$ is called a 
\emph{polynomial}. We let $S \llangle \Sigma^* \rrangle$ 
and $S\langle \Sigma^* \rangle$ respectively denote the collection of all
power series  and polynomials over $S$ in the letters $\Sigma$.
More generally, when $S' \subseteq S$, we let $S'\llangle \Sigma^* \rrangle$
denote the set of all power series in $S\llangle \Sigma^* \rrangle$ 
all of whose coefficients are in $S'$. The set of
polynomials $S'\langle \Sigma^* \rangle$ is defined in the same way.
We denote by $S'\langle \Sigma \rangle$ (no star) the collection of those 
polynomials in $S'\langle \Sigma^* \rangle$
which are linear combinations over $\Sigma$.

We define the sum $s + s'$ and product $s \cdot s'$ of two series  
$s,s' \in S\llangle\Sigma^* \rrangle$ as follows. 
For all $w \in \Sigma^*$, 
\begin{eqnarray*}
(s+s',w) &=& (s,w) + (s',w)\\
(s\cdot s',w) &=& \sum_{uu' = w}(s,u)(s',u').
\end{eqnarray*}
We may identify any element $s\in S$ with the series, in fact polynomial,
which maps $\epsilon$ to $s$  and all other elements of $\Sigma^*$ 
to $0$. In particular, $0$ and $1$ may be viewed as polynomials. 
It is well-known that equipped with the above operations and
constants,  $S\llangle \Sigma^* \rrangle$ is 
a semi\-ring which contains $S\langle \Sigma^* \rangle$ as a
subsemi\-ring.

Note that $\B\llangle \Sigma^* \rrangle$ is isomorphic to 
the semiring of languages in $\Sigma^*$, where addition corresponds 
to set union and multiplication to concatenation. An isomorphism 
maps each series in $\B\llangle \Sigma^* \rrangle$ to its support, 
and the inverse of this isomorphism maps each language $L \subseteq \Sigma^*$ 
to its \emph{characteristic series} $s$ defined by $(s,w) = 1$ if 
$w \in L$ and $(s,w) = 0$, otherwise. 

The following fact is well-known.

\begin{thm}
\label{thm-polynomials}
Given any semirings $S,S'$, any semiring morphism $h_S: S \to S'$ 
and any function $h: \Sigma \to S'$ such that 
\begin{eqnarray}
\label{eq-commute}
(sh_S)(ah) 
&=& (ah)(sh_S)
\end{eqnarray}
 for all $a \in \Sigma$ and $s \in S$,
there is a unique semiring morphism $h^\sharp : 
S\langle \Sigma^* \rangle \to S'$ which extends both $h_S$ and $h$.
\end{thm}

The condition (\ref{eq-commute}) means that for any $s \in S$ and letter $a \in \Sigma$,
$sh_S$ \emph{commutes with} $ah$. 
In particular, since $\N$ is initial, and since when $S = \N$ the condition
(\ref{eq-commute}) holds automatically, we obtain that 
any map $\Sigma\to S'$ into a semiring $S'$ extends to a unique 
semiring morphism $\N\langle \Sigma^* \rangle \to S'$, i.e., the polynomial 
semiring $\N\langle \Sigma^* \rangle$ is freely generated by $\Sigma$ 
in the variety of semirings. In the same way, $\B\langle\Sigma^*\rangle$ 
is freely generated by $\Sigma$ in the variety of idempotent semirings.
 
\section{Conway and iteration semi\-rings}
\label{sec-Conway}

In this section, we review the notions of (partial) Conway semi\-ring and 
iteration semi\-ring. The notions and facts presented here will be used 
in the freeness results.

The definition of Conway semi\-rings involves two important 
identities of regular languages. They appear implicitly in 
Conway \cite{Conway} and were first defined explicitly 
in \cite{BEmat,BEbook}. 
Partial Conway semirings appear in \cite{Espartial}.  
Recall that an \emph{ideal} of a semi\-ring $S$ is a set $I \subseteq S$ 
which contains $0$ and satisfies $I + I \subseteq I$ and 
$SI \cup IS \subseteq I$. 

This section is based on \cite{BEbook} and \cite{Espartial}.

\begin{deff}
A \emph{partial $^*$-semi\-ring} is a semi\-ring equipped with a 
partially defined star operation $ a \mapsto a^*$ 
whose domain $D(S)$ is an ideal. A \emph{ partial Conway semi\-ring} is a 
$^*$-semi\-ring $S$ satisfying the \emph{sum star} and \emph{product
star} identities:
\begin{enumerate}
\item \emph{Sum star identity}: 
\begin{eqnarray}
\label{eq-sum star}
(a+b)^* &=& a^*(ba^*)^*
\end{eqnarray}
for all $a,b \in D(S)$. 
\item \emph{Product star identity}: 
\begin{eqnarray}
\label{eq-product star}
(ab)^* &=& 1 + a(ba)^*b,
\end{eqnarray}
for all $a,b \in S$ such that $a \in D(S)$ or $b \in D(S)$.
\end{enumerate}
A \emph{$^*$-semi\-ring} is a partial $^*$-semi\-ring $S$ with $D(S) = S$, i.e., 
the star operation is totally defined. A \emph{Conway semi\-ring} is a 
partial Conway semi\-ring which is a $^*$-semi\-ring. 
Morphisms $h: S \to S'$ of (partial) $^*$-semi\-rings and (partial) 
Conway semi\-rings preserve the ideal and star operation: if 
$a \in D(S)$ then $ah \in D(S')$ and $a^*h = (ah)^*$. 
\end{deff}

Note that in any partial Conway semi\-ring $S$, 
\begin{eqnarray}
\label{eq-fixed point left}
aa^* + 1 &=& a^*\\
\label{eq-fixed point right}
a^*a + 1 &=& a^*\\
\label{eq-zero star}
0^* &=& 1
\end{eqnarray}
for all $a \in D(S)$. Moreover, for all $a,b \in S$ with $a \in D(S)$ or $b \in D(S)$,  
\begin{eqnarray}
\label{eq-simplified product star}
(ab)^*a &=& a(ba)^*.
\end{eqnarray}
It then follows that also 
\begin{eqnarray}
\label{eq-sum star 2}
(a+b)^* &=& (a^*b)^*a^*
\end{eqnarray}  
for all $a,b \in D(S)$, which can be used instead of 
(\ref{eq-sum star}) in the definition of partial Conway semi\-rings. 
By (\ref{eq-fixed point left}) and (\ref{eq-fixed point right}), 
for  any $a,b$ in a partial Conway semi\-ring $S$,
if $a \in D(S)$ then 
$a^*b$ is a solution of the equation $x =ax + b$ and $ba^*$ is a solution of 
$x= xa + b$. In particular, $a^+ = aa^* = a^*a$ is a solution of 
both $x = ax + a$ and $x = xa + a$.  



An important feature of (partial) Conway semi\-rings is that square matrices over 
Conway semi\-rings also form (partial) Conway semi\-rings. 

\begin{deff}
Suppose that $S$ is a partial Conway semi\-ring. We turn the semi\-rings $S^{k \times k}$,
$k \geq 0$ into partial $^*$-semi\-rings. Note that $D(S)^{k \times k}$, 
the collection of all $k \times k$ matrices all of whose entries 
are in $D(S)$ is an ideal of $S^{k \times k}$. The star operation
will be defined on this ideal. When $k = 0$, $S^{k \times k}$ is trivial
as is the definition of star. When $k = 1$, we use the star operation on $S$. 
Assuming that $k > 1$ we write $k = n + 1$. For a matrix 
$\left(
\begin{array}{cc}
A & B \\
C & D
\end{array}
\right) 
$ define 
\begin{eqnarray}
\label{matrix star id}
\left(
\begin{array}{cc}
A & B \\
C & D
\end{array}
\right)^*
&=&
\left(
\begin{array}{cc}
\alpha & \beta\\
\gamma & \delta
\end{array}
\right)
\end{eqnarray}
where $A\in D(S)^{n \times n}$, $B \in D(S)^{n \times 1}$, 
$C \in D(S)^{1 \times n}$, and $D \in D(S)^{1 \times 1}$, 
and where
\begin{eqnarray*}
\begin{array}{lr}
\alpha= (A + B D^* C)^*  & \beta = \alpha B D^*\\
\gamma= \delta C A^* & \delta = (D+CA^*B)^*.
\end{array}
\end{eqnarray*}
\end{deff} 

\begin{prop} 
When $S$ is a (partial) Conway semi\-ring, so is each $S^{n \times n}$. 
Moreover, the \emph{matrix star identity} (\ref{matrix star id}) 
holds for all matrices 
$\left(
\begin{array}{cc}
A & B \\
C & D
\end{array}
\right) 
$
with 
$A\in D(S)^{n \times n}$, $B \in D(S)^{n \times m}$, 
$C \in D(S)^{m \times n}$, and $D \in D(S)^{m \times m}$,
all $n,m \geq 0$.
\end{prop}

In fact, $(AB)^* = A (BA)^*B + E_n$ holds for all rectangular matrices
$A \in D(S)^{n \times m}$ and $B \in D(S)^{m \times n}$. 

For later use we note that the following \emph{permutation 
identity} holds in all (partial) Conway semi\-rings. 

\begin{prop}
When $S$ is a partial Conway semi\-ring, $A \in D(S)^{n \times n}$
and $\pi$ is an $n \times n$ permutation matrix with transpose 
$\pi^T$,  then $(\pi A\pi^T)^* = \pi A^* \pi^T$. 
\end{prop}

Following Conway \cite{Conway}, we associate 
an identity in (partial) Conway semi\-rings with each finite group. 
Let $G$ be a finite group of order $n$. Without loss of generality 
we may assume that the elements of $G$ are the integers $1,\cdots,n$.
Moreover, because the permutation identity holds in all 
(partial) Conway semi\-rings, without loss of generality we may 
fix a sequencing of the elements and assume that $1$ is the unit 
element of $G$.

\begin{deff}
We say that the \emph{group identity associated with a finite group 
$G$} of order $n$ holds in a partial Conway semi\-ring $S$ if 
\begin{eqnarray}
\label{star group id}
e_1 M_G^* u_n &=& (a_1 + \cdots + a_n)^*
\end{eqnarray}
holds, where $a_1,\cdots,a_n$ are arbitrary elements in $D(S)$, 
and where $M_G$ is the $n \times n$ matrix
whose $(i,j)$th entry is $a_{i^{-1}j}$, for all $1 \leq i,j \leq n$, and 
$e_1$ is the $1 \times n$ $0$-$1$ matrix whose first entry is $1$
and whose other entries are $0$, 
finally $u_n$ is the $n \times 1$ matrix all of whose entries are $1$.
\end{deff}

Equation (\ref{star group id}) asserts that the sum of the entries 
of the first row of $M_G^*$ is $(a_1 + \cdots + a_n)^*$.
For example, the  group identity associated with the group
of order $2$ is 
\begin{eqnarray*}
\left(\begin{array}{cc}
1 & 0
\end{array}\right)
\left(\begin{array}{cc}
a_1 & a_2\\
a_2 & a_1
\end{array}\right)^*
\left(\begin{array}{c}
1 \\
1
\end{array}\right) 
&=& 
(a_1 + a_2)^*,
\end{eqnarray*}              
which by the matrix star identity can be written as 
\begin{eqnarray*}
(a_1 + a_2 a_1^*a_2)^*(1 + a_2a_1^*) 
&=& 
(a_1 + a_2)^*.
\end{eqnarray*}
(It is known that in Conway semi\-rings, this 
identity is further equivalent to $(a^2)^*(1 + a) = a^*$.)

\begin{deff}
We say that a Conway semi\-ring $S$ is an \emph{iteration semi\-ring} if 
it satisfies all group identities. We say that a partial Conway semi\-ring
$S$ is a \emph{partial iteration semi\-ring} if it satisfies all group identities 
(\ref{star group id}) where $a_1,\cdots,a_n$ range over $D(S)$. 
A morphism of (partial) iteration semi\-rings is a (partial) Conway 
semi\-ring morphism. 
\end{deff}



We end this section by recalling from \cite{BEbook,Espartial} that power series semi\-rings 
are (partial) iteration semi\-rings. Suppose that $S$ is a semi\-ring and $\Sigma$ 
is a set, and consider the semi\-ring $S \llangle\Sigma^* \rrangle$. 
A series $s \in S\llangle\Sigma^* \rrangle$ is called \emph{proper} 
\cite{Berstel} if $(s,\epsilon) = 0$. It is clear 
that the proper series form an ideal of 
$S \llangle \Sigma^* \rrangle$. It is well-known (see e.g. \cite{Berstel}) 
that when $s$ is proper and $r$ is any series, there is a unique 
series that solves the fixed point equation $x = sx +r$, and that this  
solution is $s^*r$, where $s^*$ is the unique solution of $y = sy +1$.

\begin{prop}
For any semiring $S$, the power series semiring 
$S\llangle \Sigma^* \rrangle$, equipped with the star operation defined 
on proper series, is a partial iteration semi\-ring. 
\end{prop}

When $S$ is a $^*$-semi\-ring, it is possible to turn star into a 
total operation. Given a series $s \in S\llangle \Sigma^* \rrangle$, 
it can be written in a unique way as $s = s_0 + r$, 
where $s_0 \in S$ and $r$ is proper. Since $s_0$ is in $S$
and $S$ has a star operation, $s_0^*$ is defined. We define $s^* = (s_0^*r)^*s_0^*$,
where $(s_0^*r)^*$ is the unique solution of the equation $x = (s_0^* r) x + 1$
as before. The following fact is a special case of a more general result proved in 
\cite{BEmat,BEbook}. 

\begin{prop}
\label{prop-ext}
When $S$ is an iteration semi\-ring, so is $S \llangle \Sigma^* \rrangle$. 
\end{prop}

\section{The Kleene-Sch\"utzenberger theorem}
\label{sec-Kleene}

Let $S$ denote a semi\-ring, let $\Sigma$ denote a set, and consider the power series 
semi\-ring $S \llangle \Sigma^* \rrangle$ which is a partial iteration
semi\-ring (or an iteration semi\-ring, if $S$ is).
As usual, we identify each letter in $\Sigma$ and each 
element of $S$ with a series. We call a series $s$ in $S \llangle \Sigma^* 
\rrangle$ \emph{rational} if $s$ belongs to the least partial iteration
subsemi\-ring of $S \llangle \Sigma^* \rrangle$ containing $S \cup \Sigma$,
i.e. when $s$ is contained in the least subsemi\-ring of $S\llangle \Sigma^* \rrangle$
containing 
$S \cup \Sigma$ closed under the  star operation. 
We let $S^{\rat}\llangle \Sigma^*\rrangle$ denote the partial iteration 
semi\-ring of all rational power series in $S \llangle \Sigma^* \rrangle$.
The Kleene-Sch\"utzenberger theorem \cite{Berstel} equates rational power series with 
the power series recognizable by (weighted) automata. For later use, 
below we give a general definition of automata applicable to all
partial Conway semi\-rings, see \cite{Espartial} and \cite{BEbook}.

\begin{deff}
Let $S$ be a partial Conway semi\-ring and suppose that
$S_0$ is a subsemi\-ring of $S$ 
and $\Sigma$ is a subset of $D(S)$. An \emph{automaton in $S$ over $(S_0,\Sigma)$} 
is a triplet $\bA = (\alpha,A,\beta)$, where for some integer $n$,
$\alpha \in S_0^{1\times n}$, $\beta \in S_0^{n \times 1}$,
and $A \in (S_0\Sigma)^{n \times n}$, where $S_0\Sigma$ is the set of all
linear combinations of the elements of $\Sigma$ with coefficients 
in $S_0$. The integer $n$ is called the dimension of $\bA$. 
The \emph{behavior} of $\bA$ is $|\bA|= \alpha A^* \beta$. 
\end{deff}

Thus, when the partial Conway semi\-ring is $S \llangle \Sigma^* \rrangle$, 
where $S$ is a semiring, 
$S_0$ is $S$ and $\Sigma$ is the collection of power series corresponding to 
the letters in $\Sigma$, we obtain the usual notion of a (weighted) automaton.
We let $S^{\rec}\llangle \Sigma^* \rrangle$ denote the collection of all 
power series which are behaviors of such automata. The Kleene-Sch\"utzebreger 
theorem is:

\begin{thm}
\label{thm-Kleene}
$S^{\rat}\llangle \Sigma^* \rrangle = S^{\rec}\llangle \Sigma^* \rrangle$. 
\end{thm}

For a proof, see \cite{Espartial}. 
Below we will call an automaton (over $(S,\Sigma)$) in $S\llangle \Sigma^* \rrangle$ also  
an automaton in $S^{\rat}\llangle \Sigma^* \rrangle$. When $\bA = (\alpha,
A,\beta)$ is an automaton in $S^{\rat}\llangle \Sigma^* \rrangle$ and 
$h$ is a function $S^{\rat}\llangle \Sigma^* \rrangle \to S'$ into a 
partial iteration semiring which is a semiring morphism on $S$,
maps $\Sigma$ into $D(S')$  and 
preserves linear combinations in $S \langle \Sigma \rangle$, then 
$\bA h= ((\alpha h), (Ah), (\beta h))$ is an automaton in $S'$ 
(over $(Sh, \Sigma h)$).

For later use we also give the following result from \cite{Espartial}.

\begin{thm}
\label{thm-morphism}
Suppose that $S$ is a semi\-ring and $\Sigma$ is a set, so that 
$S^{\rat}\llangle\Sigma^* \rrangle$ is a partial iteration semi\-ring. 
Suppose that $S'$ is a partial iteration semi\-ring and $h$ is a function 
$S^{\rat}\llangle \Sigma^* \rrangle \to S'$. Then $h$ is a 
morphism of partial iteration semirings iff the restriction of $h$ 
onto $S$ is a semiring morphism, $h$ maps $\Sigma$ to $D(S')$ 
and preserves linear combinations in $S\langle \Sigma \rangle$;
moreover, $h$ preserves the behavior of automata,
so that $|\bA|h = |\bA h|$ for all automata $\bA$ in 
$S^{\rat}\llangle \Sigma^* \rrangle$.
\end{thm}

\section{The commutative identity}
\label{sec-commutative}

In the proof of our results, we will deduce the equality 
$A^*\rho = \rho B^*$ from the 
equality $A \rho = \rho B$, where $A$ is an ${m \times m}$,
$B$ is an ${n \times n}$ matrix over a partial iteration 
semi\-ring, and $\rho$ is an $m \times n$ functional
matrix. 
The commutative identity, defined below, is a generalization of the group 
identities which holds in all (partial) iteration semi\-rings.
The commutative identity allows us to infer the implication 
above, under certain conditions. 
The commutative identity was introduced for $^*$-semi\-rings 
in \cite{BEbook} but its origins in iteration theories go back to 
\cite{Es80}. See also \cite{Esgroup}. This section is rather technical 
and all proofs may be skipped at first reading.

In order to illustrate the commutative identity and its use, 
consider the following situation.
Assume that $A$, $B$ and $\rho$ are as above, but
for simplicity assume that $\rho$ as a function is 
surjective and monotone, collapsing the first $m_1$ integers to $1$, 
the next $m_2$ integers to $2$ etc. Then write 
$A$ as a block matrix $(A_{ij})_{ij}$, where each 
$A_{ij}$ is a $m_i \times m_j$ matrix
for all $i,j = 1,\cdots,n$. The condition that 
$A \rho = \rho B$ means that each row sum of any $A_{ij}$ 
is $b_{ij}$,
the $(i,j)$th entry of matrix $B$. Similarly, 
$A^* \rho = \rho B^*$ means that $A^*$ can be written 
as a matrix of blocks of size $m_i \times m_j$, 
$i,j = 1,\cdots,n$, and for each $i$ and $j$,
each row sum of the $(i,j)$th block is equal to 
the $(i,j)$th entry of $B^*$. Now assume that 
the following stronger condition holds for the matrices
$A$ and $B$: 

\emph{There exist some row matrices  
$c_{ij}$, $i,j = 1,\cdots,n$ such that each
$b_{ij}$ is the sum of the entries of $c_{i,j}$ and 
each entry of each $A_{ij}$ is a sum of certain 
entries of $c_{ij}$ such that each entry of 
$c_{ij}$ appears \emph{exactly once} as a summand in each row
of $A_{ij}$.} 

Then the commutative identity implies 
$A^* \rho = \rho B^*$. By adding $0$'s to the row matrices 
$c_{ij}$ we can make all of them size $1 \times k$, for some 
$k$, or alternatively, as we do below,  we can make each 
$c_{ij}$ size $1 \times k_i$, so that the size 
of $c_{ij}$ only depends on $i$. 
Thus the row matrices $c_{ij}$ can be arranged in the form of 
a block matrix $C = (c_{ij})_{ij}$ as below.

Before formally defining the commutative identity, we introduce some 
notation. Let $S$ be any semi\-ring and consider matrices 
$A \in S^{m \times n}$ and $B_1,\cdots,B_m\in S^{n \times p}$.
We let $A\parallel (B_1,\cdots,B_m)$ denote the matrix
in $S^{m \times p}$ whose rows are $A_1 B_1,\cdots,A_m B_m$,
where $A_1,\cdots,A_m$ are the rows of $A$. 

\begin{deff}
Suppose that $S$ is a partial $^*$-semi\-ring. We say that 
the \emph{commutative identity} holds in $S$ if for all $C \in D(S)^{n \times k}$,
$m \times n$ functional matrix $\rho$, $k \times m$ functional matrices 
$\rho_1,\cdots,\rho_m$ and $k \times n$ functional matrices 
$\tau_1,\cdots,\tau_n$ with $\rho_i \rho = \tau_{i\rho}$ for all 
$i = 1,\cdots,m$, 
\begin{eqnarray*}
((\rho C)\parallel (\rho_1,\cdots,\rho_m))^* \rho  
&=& 
\rho (C \parallel (\tau_1,\cdots,\tau_n))^*.
\end{eqnarray*} 
\end{deff}
Note that under the assumptions we have $A \rho = \rho B$ 
for the matrices $A = (\rho C) \parallel (\rho_1,\cdots,\rho_m)$
and  $B  = C \parallel (\tau_1,\cdots,\tau_n)$, and that the 
commutative identity asserts that $A^*\rho = \rho B^*$. 
   
The commutative identity has a dual which 
also holds in all (partial) iteration semirings, see 
\cite{Esgroup,Espartial}. It can be formulated as follows. 

\begin{deff}
Suppose that $S$ is a partial $^*$-semi\-ring. We say that 
the \emph{dual commutative identity} holds in $S$ if for 
all $C \in D(S)^{k \times n}$, $m \times n$ functional 
matrix $\rho$, $k \times m$ functional matrices 
$\rho_1,\cdots,\rho_m$ and $k \times n$ functional matrices 
$\tau_1,\cdots,\tau_n$ with $\rho_i \rho = \tau_{i\rho}$ for all 
$i = 1,\cdots,m$, 
\begin{eqnarray*}
\rho^T ((\rho_1^T,\cdots,\rho_m^T) \parallel (C \rho^T))^*  
&=& 
 ( (\tau_1^T,\cdots,\tau_n^T) \parallel C)^* \rho^T.
\end{eqnarray*} 
\end{deff}

Here $(B_1,\cdots,B_n)\parallel A$ is 
the matrix whose columns are $B_1 A_1,\cdots,B_n A_n$,
where $A_1,\cdots,A_n$ are the columns of $A$,
$A$ is $m \times n$ and $B_1,\cdots,B_n$ are $p \times m$.

\begin{deff}
A semi\-ring $S$ is \emph{atomistic} if 
for any $a_1,\ldots,a_m$ and $b_1,\cdots,b_n$ in $S$ if 
$a_1+ \cdots + a_m = b_1 + \cdots + b_n$ then there 
exist $c_1,\cdots,c_k$ in $S$ and \emph{partitions} $I_1,\cdots,I_m$
and $J_1,\cdots,J_n$ of the set $\{1,\cdots,k\}$ such that 
\begin{eqnarray*}
a_i &=& \sum_{p \in I_i}c_p\\
b_j &=& \sum_{q \in J_j}c_q
\end{eqnarray*}
for each $i= 1,\cdots,m$ and $j = 1,\cdots,n$. 
\end{deff}

Examples of atomistic semi\-rings are $\B$, $\N$ and  $\N_\infty$.

\begin{prop}
\label{prop-reducible}
Suppose that $S$ is atomistic and $A \in S ^{m \times m}$, 
$B \in S^{n \times n}$ are such that 
$A \rho = \rho B$ holds for some $m \times n$ functional 
matrix $\rho$. Then there is a matrix $C \in S^{n \times k}$,
 $k \times m$ functional matrices 
$\rho_1,\cdots,\rho_m$ and $k \times n$ functional matrices 
$\tau_1,\cdots,\tau_n$ with $\rho_i \rho = \tau_{i\rho}$ for all 
$i = 1,\cdots,m$ such that  
\begin{eqnarray*}
A &=& (\rho C)\parallel (\rho_1,\cdots,\rho_m)\\ 
B &=&  C \parallel (\tau_1,\cdots,\tau_n).
\end{eqnarray*} 
\end{prop}

{\sl Proof.} It suffices to consider the case when $\rho$ is surjective and monotone. 
Thus, the assumption is that $A$ is a block matrix $(A_{ij})_{ij}$
such that the sum of each row of each $A_{ij}$ is $b_{ij}$, the 
$(i,j)$th entry of $B$. Since $S$ is atomistic, 
for each $(i,j)$ there is 
a row matrix $C_{ij}$ such that the sum of its entries is $b_{ij}$ 
and each entry of each row of each $A_{ij}$ can be written as a sum
of certain entries of $C_{ij}$ in such a way that each entry of 
$C_{ij}$ appears exactly once as a summand in each row of $A_{ij}$.
But this is clear since the semiring is atomistic. \eop 

In a similar way, we have:

\begin{prop}
\label{prop-reducible-T}
Suppose that $S$ is atomistic and $A \in S ^{m \times m}$, 
$B \in S^{n \times n}$ are matrices such that 
$ \rho^T A= B \rho^T$ holds for some $m \times n$ functional 
matrix $\rho$. Then there is a matrix $C \in S^{k \times n}$,
 $k \times m$ functional matrices 
$\rho_1,\cdots,\rho_m$ and $k \times n$ functional matrices 
$\tau_1,\cdots,\tau_n$ with $\rho_i \rho = \tau_{i\rho}$ for all 
$i = 1,\cdots,m$ such that  
\begin{eqnarray*}
A &=& (\rho_1^T,\cdots,\rho_m^T)\parallel (C\rho^T)\\ 
B &=& (\tau_1^T,\cdots,\tau_n^T) \parallel C.
\end{eqnarray*} 
\end{prop}

\begin{prop}
\label{prop-reducible2}
Suppose that $S$ is atomistic. Let 
$A \in S \langle \Sigma \rangle^{m \times m}$, 
$B \in S \langle \Sigma \rangle^{n \times n}$
be matrices and let $\rho$ be a functional matrix of size $m \times n$. 
If $A \rho = \rho B$ then there is a matrix $C \in S \langle \Sigma \rangle^{n \times k}$,
 $k \times m$ functional matrices 
$\rho_1,\cdots,\rho_m$ and $k \times n$ functional matrices 
$\tau_1,\cdots,\tau_n$ with $\rho_i \rho = \tau_{i\rho}$ for all 
$i = 1,\cdots,m$ such that  
\begin{eqnarray*}
A &=& (\rho C)\parallel (\rho_1,\cdots,\rho_m)\\ 
B &=&  C \parallel (\tau_1,\cdots,\tau_n).
\end{eqnarray*} 
\end{prop}

{\sl Proof.}
There exists a finite $\Sigma_0 \subseteq\Sigma$ such that 
whenever $A$ or $B$ has an entry which has a summand $s\sigma$ 
where $\sigma \in \Sigma$ and $s$ is \emph{not} $0$, then $\sigma \in \Sigma_0$.
Now for each $\sigma \in \Sigma_0$, let $A^\sigma$ denote the
$m \times m$  matrix whose $(i,j)$th entry for $i,j = 1,\cdots,m$ is $s\sigma$ 
where $s$ is the coefficient of $\sigma$ in $A_{ij}$, the $(i,j)$th 
entry of $A$. If there is no such summand, let $A_{ij}^\sigma = 0$.
Define the $n \times n$ matrices $B^\sigma$, $\sigma \in \Sigma_0$ 
in the same way. We then have $A^\sigma\rho = \rho B^\sigma$, 
for each $\sigma \in \Sigma_0$. Thus, by Proposition~\ref{prop-reducible},
for each $\sigma \in \Sigma_0$ there is a matrix 
$C^\sigma \in S\langle \{\sigma\} \rangle^{m \times k^\sigma}$
and functional matrices $\rho_i^\sigma$ and $\tau_j^\sigma$,
$i = 1,\cdots,m$, $j = 1,\cdots,n$ of 
appropriate size with $\rho_i^\sigma \rho = \tau_{i\rho}^\sigma$ such that 
\begin{eqnarray*}
A^\sigma &=& (\rho C^\sigma)\parallel (\rho_1^\sigma,\cdots,\rho_m^\sigma)\\ 
B^\sigma &=&  C^\sigma \parallel (\tau_1^\sigma,\cdots,\tau_n^\sigma).
\end{eqnarray*}
Let $\Sigma_0 = \{\sigma_1,\cdots,\sigma_p\}$, say. Define 
\begin{eqnarray*}
C &=& 
\left(\begin{array}{ccc}
C^{\sigma_1} & \cdots &C^{\sigma_p}
\end{array}\right)
\end{eqnarray*}
and 
\begin{eqnarray*}
\rho_i &=& 
\left(
\begin{array}{c}
\rho_i^{\sigma_1}\\
\vdots\\
\rho_i^{\sigma_p}
\end{array}
\right)\\
\tau_j &=& 
\left(
\begin{array}{c}
\tau_j^{\sigma_1}\\
\vdots\\
\tau_j^{\sigma_p}
\end{array}
\right)
\end{eqnarray*}
for all $i = 1,\cdots,m$ and $j = 1,\cdots,n$. 
Then for each $i$, 
\begin{eqnarray*}
\rho_i \rho &=&
\left(
\begin{array}{c}
\rho_i^{\sigma_1}\\
\vdots\\
\rho_i^{\sigma_p}
\end{array}
\right)\rho \\
&=& 
\left(
\begin{array}{c}
\rho_i^{\sigma_1}\rho\\
\vdots\\
\rho_i^{\sigma_p}\rho
\end{array}
\right)\\
&=&
\left(
\begin{array}{c}
\tau_{i\rho}^{\sigma_1}\\
\vdots\\
\tau_{i\rho}^{\sigma_p}
\end{array}
\right)\\
&=& 
\tau_{i\rho}.
\end{eqnarray*}
Also,
\begin{eqnarray*}
(\rho C)\parallel (\rho_1,\cdots,\rho_m)
&=& 
\left(
\begin{array}{c}
C_{1\rho}^{\sigma_1}\rho_1^{\sigma_1}+ \cdots + C_{1\rho}^{\sigma_p} \rho_1^{\sigma_1}\\
\vdots\\
C_{m\rho}^{\sigma_1}\rho_1^{\sigma_m}+ \cdots + C_{m\rho}^{\sigma_p} \rho_m^{\sigma_1}
\end{array}
\right)\\
&=& 
\left(
\begin{array}{c}
A_1^{\sigma_1} + \cdots + A_1^{\sigma_p}\\
\vdots\\
A_m^{\sigma_1} + \cdots + A_m^{\sigma_p}
\end{array}
\right)\\
&=& 
A,
\end{eqnarray*}
where $A_1,\cdots,A_m$ denote the rows of $A$.
In a similar way, 
$C \parallel (\tau_1,\cdots,\tau_n) = B$. \eop 

Symmetrically, we have:

\begin{prop}
\label{prop-reducuble2-T}
Suppose that $S$ is atomistic. Let
$A \in S \langle \Sigma \rangle^{m \times m}$, 
$B \in S \langle \Sigma \rangle^{n \times n}$
and let $\rho$ be a functional matrix of size $m \times n$. 
If $\rho^T A  = B \rho^T$ then there is a matrix 
$C \in S \langle \Sigma \rangle^{k \times n}$,
 $k \times m$ functional matrices 
$\rho_1,\cdots,\rho_m$ and $k \times n$ functional matrices 
$\tau_1,\cdots,\tau_n$ with $\rho_i \rho = \tau_{i\rho}$ for all 
$i = 1,\cdots,m$ such that  
\begin{eqnarray*}
A &=& (\rho_1^T,\cdots,\rho_m^T)\parallel (C\rho^T)\\ 
B &=& (\tau_1^T,\cdots,\tau_n^T) \parallel C.
\end{eqnarray*} 
\end{prop}

\section{Free partial iteration semi\-rings}
\label{sec-Nrat}

In this section, our aim is to show that 
for any set $\Sigma$, 
$\N^{\rat} \llangle \Sigma^* \rrangle$
is freely generated by $\Sigma$ in the 
class of \emph{partial} iteration semi\-rings. 
For this reason, assume that $S$ is a 
partial iteration semi\-ring and $h$ is a 
function $\Sigma \to D(S)$. We can 
extend $h$ to a semi\-ring morphism 
$\N \langle \Sigma^* \rangle \to S$.
In particular, $h$ is defined on $\N$ and 
on $\N\langle \Sigma \rangle$, and in a pointwise manner,
on matrices with entries in $\N$ or $\N\langle \Sigma \rangle$.

We want to show that $h$ can be extended 
to a unique morphism $h^\sharp : 
\N^{\rat} \llangle \Sigma^* \rrangle \to S$ 
of partial iteration semi\-rings. For this reason,
we will consider automata $\bA = (\alpha,A,\beta)$  
(in $\N^{\rat}\llangle \Sigma^* \rrangle$) where 
$\alpha \in \N^{1 \times n}$,
$\beta \in \N^{n \times 1}$ and 
$A \in \N\langle \Sigma \rangle^{n \times n}$ 
for some $n$. Using the function $h$, we define the 
\emph{image of $\bA$} as the automaton   
$\bA h$ in $S$: $\bA h = (\alpha h, Ah, \beta h)$.
We know from Theorem~\ref{thm-morphism} that we are forced to define $h^\sharp$ by 
$|\bA|h = |\bA h|$, for all automata $\bA$. 
We also know that if this function is well-defined, then it 
is a morphism $\N\llangle \Sigma^* \rrangle \to S$ 
of partial iteration semi\-rings (which clearly extends $h$). 
So all we have to show is that $h^\sharp$ is well-defined. 
The proof of this fact relies on a result proved in \cite{Sakarovitchetal2}
that we recall now.

\begin{deff}
Let $\bA = (\alpha,A,\beta)$ and $\bB = (\gamma,B,\delta)$ be two automata
(in $\N^{\rat}\llangle \Sigma^* \rrangle$) 
of dimension $m$ and $n$, respectively. We say that an $m \times n$ functional 
matrix $\rho$ is a \emph{simulation} $\bA \to \bB$ if $\alpha \rho = \gamma$, 
$\rho \delta = \beta$ and $A \rho = \rho B$ hold. Moreover, we say that 
$\rho$ is a \emph{dual simulation} $\bA \to \bB$ if $\rho$ is a simulation 
$\bA^T \to \bB^T$, where $\bA^T = (\beta^T, A^T,\alpha^T)$ and 
$\bB^T$ is defined in the same way.
\end{deff}
Note that $\rho$ is a dual simulation $\bA \to \bB$ iff $\gamma \rho^T = \alpha$,
$B\rho^T = \rho^T A$ and $\rho^T\beta = \delta$ hold. 

(More general simulations were defined in \cite{BEbook}.
The simulations defined above are the functional and dual functional 
simulations of \cite{BEbook}. 
In the papers \cite{Sakarovitchetal1,Sakarovitchetal2}, the terms ``covering''
and ``co-covering'' are used for simulation and dual simulation. Moreover, 
only simulations and dual simulations corresponding to
surjective functions are considered, since in the formulation of 
Theorem~\ref{thm-Sakarovitchetal} given in \cite{Sakarovitchetal2},
the automata are ``trim'', i.e., without useless states.) 
Let $\sim$ denote the least equivalence relation such that $\bA \sim \bB$ 
holds whenever there is a functional simulation or a dual functional simulation
$\bA \to \bB$. Moreover, call two automata $\bA$ and $\bB$ 
\emph{equivalent} if $|\bA| = |\bB|$. The following result 
was proved in \cite{Sakarovitchetal2}:

\begin{thm}
\label{thm-Sakarovitchetal}
Two automata $\bA$ and $\bB$ in $\N^{\rat}\llangle \Sigma^* \rrangle$ 
are equivalent iff $\bA \sim \bB$. 
\end{thm}

So  our task reduces to showing that for automata 
$\bA$ and $\bB$ in $\N^{\rat}\llangle \Sigma^* \rrangle$,
if there is a functional or a dual functional simulation
$\bA \to \bB$, then $|\bA h| =|\bB h|$. 

\begin{lem}
\label{lem-unique}
Suppose that $\bA = (\alpha, A,\beta)$ and 
$\bB = (\gamma, B,\delta)$ are  automata 
in $\N^{\rat}\llangle \Sigma^*  \rrangle$ of 
dimension $m$ and $n$, respectively.
Suppose that $\rho$  is an $m \times n$  
functional matrix  which is a simulation $\bA \to \bB$.
Then $|\bA  h| = |\bB h|$.
\end{lem}

{\sl Proof.} Since $A \rho =\rho B$, it follows from Proposition~\ref{prop-reducible2} 
that there exists a matrix $C \in \N \langle \Sigma \rangle^{n \times k}$ 
and $k \times m$ functional matrices $\rho_1,\cdots,\rho_m$
and $k \times n$ functional matrices $\tau_1,\cdots,\tau_n$
with $\rho_i \rho = \tau_{i\rho}$ for all $i$
such that $A = (\rho C ) \parallel (\rho_1,\cdots,\rho_m)$
and $B = C \parallel (\tau_1,\cdots,\tau_n)$.
Thus, also $Ah = ((\rho h)(Ch) ) \parallel (\rho_1h,\cdots,\rho_mh)$
and $Bh = (Ch) \parallel (\tau_1h,\cdots,\tau_nh)$.
 Thus, by the commutative identity,
$(Ah)^*(\rho h) = (\rho h) (Bh)^*$. Thus, 
\begin{eqnarray*}
|\bA h|
&=& (\alpha h) (Ah)^* (\beta h)   \\
&=& (\alpha h) (Ah)^* ((\rho \delta)h)\\
&=& (\alpha h) (Ah)^* (\rho h) (\delta h)\\
&=& (\alpha h) (\rho h) (Bh)^* (\delta h)\\
&=& ((\alpha \rho)h) (Bh)^* (\delta h)\\
&=& (\gamma h) (Bh)^* (\delta h)\\
&=& |\bB h|. \eop
\end{eqnarray*}

\begin{lem}
\label{lem-unique2}
Suppose that $\bA$ and 
$\bB$ are finite automata as above of dimension $m$ and $n$, respectively.
Suppose that $\rho$ is an $m \times n$ functional matrix 
which is a dual simulation 
$\bA \to \bB$. Then $|\bA h| = |\bB h|$.
\end{lem}

{\sl Proof.} Since $\rho$ is a simulation $\bA^T \to \bB^T$,
it follows as above that 
$A^T = (\rho C^T ) \parallel (\rho_1,\cdots,\rho_m)$
and $B^T = C^T \parallel (\tau_1,\cdots,\tau_n)$ for 
some $C^T$, $\rho_1,\cdots,\rho_m$ and $\tau_1,\cdots,\tau_n$
with $\rho_i \rho = \tau_{i\rho}$.
Thus, 
$A = (\rho_1^T,\cdots,\rho_m^T) \parallel (C\rho^T)$ and 
$B = (\tau_1^T,\cdots,\tau_n^T)\parallel C$. 
The proof can be completed as above using the 
dual commutative identity. \eop 

The main result of this section is: 

\begin{thm}
\label{thm-free partial}
$\N^{\rat}\llangle \Sigma^* \rrangle$ is freely generated by 
$\Sigma$ in the class of partial iteration semi\-rings. 
In detail, given any partial iteration semi\-ring $S$ and function 
$h: \Sigma \to D(S)$, there is a unique partial 
iteration semi\-ring morphism $h^\sharp : 
\N^{\rat} \llangle \Sigma^* \rrangle \to S$ extending $h$.
\end{thm}

{\sl Proof.}
Given $S$ and $h$, define $h^\sharp$ as follows. 
First, extend $h$ to a  semiring morphism 
$\N\langle \Sigma^* \rangle \to S'$. By 
Theorem~\ref{thm-Kleene}, we know that every rational series in 
$\N^{\rat}\llangle \Sigma^* \rrangle$ is the behavior of an automaton 
in $\N^{\rat}\llangle \Sigma^* \rrangle$. We also know that for any 
rational power series $r \in \N^{\rat} \llangle \Sigma^* \rrangle$ recognized by an 
automaton $\bA$, we are forced to define $rh^\sharp  = 
|\bA h|$. By Theorem~\ref{thm-Sakarovitchetal}, Lemma~\ref{lem-unique} and 
Lemma~\ref{lem-unique2},
$h^\sharp$ is well-defined. It is clear that $h^\sharp$ extends $h$.
Moreover, by Theorem~\ref{thm-morphism}, $h^\sharp$ is a morphism
of partial iteration semi\-rings. 
\eop


\begin{remark}
The paper \cite{Espartial} also defines \emph{partial iterative 
semirings} as partial $^*$-semirings $S$ such that for each
$a,b \in S$, if $a \in D(S)$, then $a^*b$ is the \emph{unique} 
solution of the equation $x = ax + b$. It is shown that every 
partial iterative semiring is a partial iteration semiring,
and that for any semiring $S$ and set $\Sigma$, the power series 
semiring $S\llangle \Sigma^* \rrangle$ is a partial iterative 
semiring. Thus, $S^{\rat}\llangle \Sigma^* \rrangle$ is also 
a partial iterative semiring. Since morphisms of partial iterative 
semirings preserve star, it follows that $\N^{\rat}\llangle \Sigma^* \rrangle$ 
is the free partial iterative semiring on $\Sigma$. 
This fact is related to a result proved in \cite{MorisakiSakai},
where Morisaki and Sakai extended Salomaa's axiomatization 
\cite{Salomaa} of regular languages to rational power series over \emph{fields}
(or more generally, \emph{principal ideal  domains}). 
\end{remark}

Theorem~\ref{thm-free partial} can be generalized. 
Consider a power series semiring 
$S^\rat \llangle \Sigma^* \rrangle$ where $S$ is any semiring.
We can define simulations and dual simulations and the relation 
$\sim$  for automata in $S^{\rat}\llangle \Sigma^* \rrangle$ 
over $(S,\Sigma)$ in the same way as above.
For example, when $\bA = (\alpha,A,\beta)$ and $\bB = (\gamma,B,\delta)$ 
are automata over $(S,\Sigma)$ of dimension $m$ and $n$, then a 
simulation $\bA \to \bB$ is an $m \times n$ functional matrix 
$\rho$ such that $\alpha \rho = \gamma$, $A\rho = \rho B$ and 
$\beta = \rho \delta$. If $\rho$ is a simulation $\bA \to \bB$, then 
$$\alpha A^k \beta = \alpha A^k \rho \delta = \alpha \rho B^k \delta = 
\gamma B^k \delta $$ 
for all $k$, and thus $|\bA| = |\bB|$, i.e.,
$\bA$ and $\bB$ are equivalent. In a similar way, if $\rho$ is a dual 
simulation $\bA \to \bB$, then $|\bA| = |\bB|$. Thus, if $\bA \sim \bB$, then 
$\bA$ and $\bB$ are equivalent. In the following generalization of 
Theorem~\ref{thm-free partial} we will assume that also the converse
property is true, if $\bA$ and $\bB$ are equivalent then 
$\bA \sim \bB$.

\begin{thm}
\label{thm-gen}
Let $S$ be a semiring and $\Sigma$ a set. 
Suppose that if two automata $\bA,\bB$ in $S^\rat\llangle \Sigma^* \rrangle$ 
over $(S,\Sigma)$ are equivalent then $\bA \sim \bB$ holds. Moreover,
suppose that $S$ is atomistic. Then $S^\rat\llangle \Sigma^* \rrangle$ 
has the following universal property. Given any partial iteration semiring 
$S'$, semiring morphism $h_S: S \to S'$ and function $h: \Sigma \to S'$
such that $sh_S$ commutes with $ah$ for all $s \in S$ and $a \in \Sigma$, 
there is a unique partial iteration semiring morphism 
$S^\rat\llangle \Sigma^* \rrangle\to S'$ extending $h_S$ and $h$. 
\end{thm}

The proof is exactly the same. Theorem~\ref{thm-gen} is applicable 
for example to the boolean semiring $\B$ (see \cite{BEbook}),
and the semirings $\k$ defined in Section~\ref{sec-Ninftyrat1}.  
However, for rings simpler characterizations exist, cf. \cite{Berstel}.

\begin{remark}
Without the  assumption that $S$ is atomistic,
we only have the following fact. Suppose that $S'$ is a partial 
Conway semiring satisfying the \emph{functorial star implications} 
\cite{BEbook} 
\begin{eqnarray*}
A \rho =\rho a &\Rightarrow & A^* \rho = \rho a^*\\
\rho^T A = a \rho^T &\Rightarrow & \rho^T A^* = a^* \rho^T
\end{eqnarray*} 
for all $a \in D(S')$ and $A \in S'^{m \times m}$ whose entries are 
in $D(S')$, and for all $m \times 1$ functional matrices, 
$m \geq 2$. Then, as shown in 
\cite{BEbook,Espartial},  $S'$ is a 
partial iteration semiring satisfying the functorial star implications
\begin{eqnarray*}
A \rho =\rho B &\Rightarrow & A^* \rho = \rho B^*\\
\rho^T A = B \rho^T &\Rightarrow & \rho^T A^* = B^* \rho^T
\end{eqnarray*} 
for all  $A \in S'^{m \times m}$, $B \in S'^{n \times n}$
whose entries are in $D(S')$, 
and for all $m \times n$ functional matrices $\rho$ for any integers $m,n$.
As above, suppose that if two automata $\bA,\bB$ in $S^\rat\llangle \Sigma^* \rrangle$ 
over $(S,\Sigma)$ are equivalent then $\bA \sim \bB$.
Then for any semiring morphism $h_S: S \to S'$ and function $h: \Sigma \to S'$
such that $\Sigma h \subseteq D(S')$
and $sh_S$ commutes with $ah$, for all $s \in S$ and $a \in \Sigma$,
there is a unique partial iteration semiring morphism 
$S^\rat\llangle \Sigma^* \rrangle\to S'$ extending $h_S$ and $h$.
\end{remark}

\section{A characterization}
\label{sec-Ninftyrat1}

We have seen that for any set $\Sigma$, $\N^{\rat} \llangle \Sigma^* \rrangle$
is freely generated by $\Sigma$ in the class of \emph{partial} iteration
semi\-rings. In particular, $\N$ is initial in the class of 
partial iteration semi\-rings. This latter fact is also clear by noting that  
the star operation is completely undefined in $\N$ and that $\N$ is initial 
in the class of semi\-rings. The smallest iteration semi\-ring which contains
$\N$ as a subsemi\-ring is $\N_\infty$, the completion of $\N$ with a point 
of infinity denoted $\infty$ and star operation defined by $0^* =1$ 
and $n^* = \infty$  for all $n \neq 0$. 
In this section our aim is to show that the iteration semi\-rings
$\N_\infty^{\rat} \llangle \Sigma^* \rrangle$ are the free algebras 
in a subvariety of iteration semi\-rings defined by a few simple identities. 
By Proposition~\ref{prop-ext}, $\N_\infty^{\rat} \llangle \Sigma^* \rrangle$ is 
an iteration semiring.

The structure of the initial iteration semi\-ring was described in 
\cite{BEbook}. Its elements are 
$$0,1,2,\cdots,1^*,(1^*)^2,\cdots,1^{**},$$
ordered as indicated. Sum and product on the integers are the standard 
operations; the sum and product on the remaining 
elements are given by:
\begin{eqnarray*}
	x + y & = & \max \{ x, y \},\quad \hbox{if } x \geq 1^* 
                     \hbox{ or } y  \geq 1^*     \\
   (1^*)^n(1^*)^p & = & (1^*)^{n+p} \\
   x1^{**}  =  1^{**}x &=& 1^{**},\quad \hbox{if } x \neq 0     .
\end{eqnarray*}
Lastly, the star operation is defined by:
\begin{eqnarray*}
x^*  & = &  \left\{\begin{array}{ll} 
1     &  \hbox{ if } x =  0  \\ 
1^*     &  \hbox{ if } x = 1  \\
1^{**} & \hbox{ otherwise.}
\end{array} 
\right.  
\end{eqnarray*}
Identifying $1^*$ and $1^{**}$, the resulting congruence collapses 
the elements $$1^*,(1^*)^2,\cdots, 1^{**},$$ so that the corresponding 
quotient $^*$-semi\-ring is isomorphic to $\N_\infty$.

In this section we will characterize the iteration semi\-rings 
$\N_\infty^{\rat} \llangle \Sigma^* \rrangle$ as the free algebras in 
the subvariety $\V$ of iteration semi\-rings 
specified by the following identities:
\begin{eqnarray}
\label{ax1}
1^*1^* &=& 1^*\\
\label{ax2}
1^* a &=& a 1^*\\
\label{ax3}
1^*(1^*a)^* &=& 1^*a^*.
\end{eqnarray}

\begin{prop}
The identity $1^* = 1^{**}$ holds in $\V$. 
\end{prop}

{\sl Proof.} Instantiating (\ref{ax3}) with $a = 1$ and using 
(\ref{ax1}) we have $1^*1^{**} = 1^*1^* = 1^*$. 
But by the above description of the initial iteration semi\-ring, 
$1^*1^{**} = 1^{**}$ in any iteration semi\-ring. \eop 

Since $2^* = 3^* = \cdots = 1^{**}$ in the initial iteration semiring,
it follows that $n^* = 1^*$ holds in $\V$ for any integer $n \geq 1$
viewed as a term. 

Thus, (\ref{ax1}) may be replaced by the identity $1^* = 1^{**}$.
Also, by (\ref{eq-sum star 2}), $(1 + a)^* = (1^*a)^*1^*$, 
so that in view of 
(\ref{ax2}), equation (\ref{ax3}) is equivalent to 
\begin{eqnarray}
(1 + a)^* &=&  1^*a^*.
\end{eqnarray}
More generally, we have that 
\begin{eqnarray}
\label{n plus a}
(n + a)^* &=& 1^*a^*
\end{eqnarray}
holds in $\V$, for any $n \in \N_\infty$, $n \neq 0$ viewed as a term.
Also, in view of the other axioms, (\ref{ax3}) is equivalent to the simpler
\begin{eqnarray}
\label{ax3-simpler}
a^{**} = 1^*a^*
\end{eqnarray} 
since  $(a+1)^* = 1^*(a1^*)^* = 1^* + 1^*(a1^*)^+ = 
1 +  1^* + 1^*(a1^*)^+ = 1 + (a+1)^* = a^*a^{**} + 1 = a^{**}$
using only the Conway identities.

Since $\N_\infty$ satisfies the identities (\ref{ax1}), 
(\ref{ax2}) and (\ref{ax3}), and since $1^* = 1^{**}$ holds in $\V$, 
we have:

\begin{cor}
\label{cor-initial}
$\N_\infty$ is initial in $\V$.
\end{cor}

Also, for each set $\Sigma$, both $\N_\infty\llangle \Sigma^* \rrangle$ 
and $\N_\infty^{\rat}\llangle \Sigma^* \rrangle$ are in $\V$.

Consider the set of \emph{iteration semiring terms}, or just {\em terms} 
over $\Sigma$ defined by
\begin{eqnarray*}
t &=& 0\ |\ 1\ |\ a,\ a \in \Sigma\ |\ t+t\ |\ t\cdot t\ |\ t^*\ .
\end{eqnarray*}
A term is called \emph{constant term} or just \emph{constant}  
if it contains no
occurrence of any letter in $\Sigma$.
We will say that two terms $s,t$ are 
\emph{equivalent}   
if they are equivalent modulo the defining identities of $\V$, i.e., 
when the identity $s = t$ \emph{holds} in $\V$.

Each term $t$ over $\Sigma$ evaluates to a series $|t|$ in 
$\N_\infty^{\rat}\llangle \Sigma^* \rrangle$ as usual.   
Since $\N_\infty$ is initial in $\V$, for any constant terms 
$s,t$ we have $|s| = |t|$ iff $s = t$ holds in $\V$. 
We may thus identify each constant term with 
an element of $\N_\infty$.

The class $I$ of \emph{ideal terms} is the least class of terms with
the following properties. 
\begin{enumerate}
\item $0 \in I$ and $a \in I$ for all $a \in \Sigma$.
\item If $s \in I$ and $t \in I$ then $s + t \in I$.
\item If $s \in I$ and $t \in I$ or $t$ is a constant
in $\N$, then $st$ and $ts$ are in $I$.
\item If $s \in I$ then $s^+$ is in $I$, where $s^+$   is an abbreviation 
for $ss^*$.
\end{enumerate} 

\begin{lem}
When $t$ is ideal, $|t|$ is proper and $|t| \in \N\llangle \Sigma^* \rrangle$.  
\end{lem}

The easy proof is omitted. 
It then follows that each ideal term $t$ also evaluates to a series in the 
partial iteration semiring $\N^{\rat}\llangle \Sigma^* \rrangle$, and that 
this series is the same as the evaluation of $t$ in 
$\N_\infty^{\rat}\llangle \Sigma^* \rrangle$.

\begin{lem}
\label{lem-3parts1}
For every term $t$ there is an equivalent term of the form $t_c + t_0 + 1^*t_\infty$, 
where $t_c$ is a constant in $\N$, $t_0$ is an ideal term, and $t_\infty$ 
is a term. Moreover, if $t_c \neq 0$ then $t_\infty$ is ideal. 
\end{lem}

{\sl Proof.} This fact is implied by the following claim:
 
\emph{For every term $t$ there is an equivalent term of the form $t_c + t_0 + 1^*t_\infty$, 
where $t_c$ is a constant in $\N_\infty$, $t_0$ and $t_\infty$ are ideal terms.}

We prove this fact by induction on the structure of $t$. 
When $t$ is $0,1$ or a letter in $\Sigma$, our claim is clear.
Suppose that $t = p + s$. Then $t$ is equivalent to 
$(p_c + s_c) + (p_0 + s_0) + 1^*(p_\infty + s_\infty)$.
Assume now that $t = ps$. Then using (\ref{ax1}) 
and (\ref{ax2}), $t$ is equivalent to 
$p_cs_c + (p_cs_0 + p_0s_c) + 1^*((p_c +p_0)s_\infty + p_\infty(s_c + s_0) 
+ p_\infty s_\infty)$. 
  Finally, assume that $t = s^*$. If $s_c = 0$, then 
$t$ is equivalent to $1 + s_0^+ + 1^*(s_0 + s_\infty)^*s_\infty s_0^*$ 
as shown by the following computation using the sum star 
and product star identities and (\ref{ax1}), (\ref{ax2}) 
and (\ref{ax3}).
\begin{eqnarray*}
(s_0 + 1^*s_\infty)^* 
&=& 
s_0^*(1^*s_\infty s_0^*)^*\\
&=& 
s_0^*(1 + 1^*(s_\infty s_0^* 1^*)^*s_\infty s_0^*)\\
&=& 
s_0^* + s_0^*1^*(1^* s_\infty s_0^*)^* s_\infty s_0^*\\
&=& 
s_0^* + s_0^*1^*(s_\infty s_0^*)^* s_\infty s_0^*\\
&=& 
s_0^* + 1^* s_0^* (s_\infty s_0^*)^*s_\infty s_0^*\\
&=&
1 + s_0^+ + 1^*(s_0 + s_\infty)^*s_\infty s_0^*.
\end{eqnarray*}
If $s_c \neq 0$, then using (\ref{n plus a}) 
we have that $t$ is equivalent to $1^*(s_0 + s_\infty)^* = 1^* + 1^*(s_0 + s_\infty)^+$.
In either case, $s^*$ is of the required form. 
\eop

As an immediate corollary, we note the following Fatou property: 

\begin{cor}
If $s \in \N_\infty^{\rat} \llangle \Sigma^* \rrangle$ and all coefficients 
of $s$ are in $\N$, then $s \in \N^{\rat}\llangle \Sigma^* \rrangle$.
\end{cor}

In our proof that each iteration semiring 
$\N_\infty^{\rat} \llangle \Sigma^* \rrangle$ 
is freely generated by $\Sigma$ in the variety $\V$ we will make
use of the corresponding fact for the boolean semiring,
proved in Krob \cite{Krob}.

\begin{thm}
\label{thm-Krob}
For each $\Sigma$, $\B^{\rat}\llangle \Sigma^* \rrangle$ is
freely generated by $\Sigma$ in the variety of all iteration 
semirings satisfying $1^* = 1$.
\end{thm}

Let $\W$ denote the variety of iteration semirings satisfying $1^* = 1$.
It is clear that $\W$ is a subvariety of $\V$. We introduce a 
construction which assigns to every iteration semiring $A$ in $\V$
an iteration semiring $1^*A$ in $\W$.

Suppose that $A \in \V$. We define $1^*A = \{1^*a : a \in A\}$.
It is clear that $1^*A$ contains $0$ and is closed under sum
and product. Also, using (\ref{ax1}), (\ref{ax2}) and (\ref{ax3}),
$(1^*a)^+ = 1^*a(1^*a)^* = 1^*a1^*a^* = 1^* a^+$, showing that 
$1^*A$ is closed under the ``plus operation'' $a \mapsto a^+$. However, $1^*A$ does 
not necessarily contain $1$ and is not necessarily closed under star.

\begin{deff}
For each $A \in \V$, we equip $1^*A$ with the following operations 
and constants. The sum $+$  and product $\cdot$ operations 
and the constant $0$ are inherited from $A$, the constant $\b1$ is $1^*$ 
and the star operation $^\otimes$ is defined by $(1^*a)^\otimes 
= 1^*(1^*a)^*$.
\end{deff}

Note that $^\otimes$ is well-defined, since if $1^*a = 1^*b$,
for some $a,b \in A$, then $1^*(1^*a)^* = 1^*(1^*b)^*$.
Also, by (\ref{ax3}), $(1^*a)^\otimes = 1^*a^*$. 
Using this, it follows that the plus operation of $1^*A$ determined by 
the star operation $^\otimes$  is the restriction of the
plus operation of $A$. Indeed, for all $a \in A$,
$1^*a(1^*a)^\otimes = 1^*a1^*a^*= 1^*a^+ = (1^*a)^+$.

\begin{lem}
\label{lem-VW}
For any $A \in \V$, the assignment 
$h: a \mapsto 1^*a$, $a \in A$  preserves all
operations and constants. 
\end{lem}

{\sl Proof.} Clearly, we have $0h = 0$ and $1h = \b1$. 
Also, $(a+b)h = 1^*(a+b) =1^*a + 1^*b = ah + bh$
and $(ab)h = 1^*ab = 1^*1^*ab = 1^*a1^*b = (ah)(bh)$,
for all $a,b \in A$. Finally, $a^*h = 
1^*a^* = (1^*a)^\otimes = (ah)^\otimes$. \eop

\begin{cor}
\label{cor-VW}
For each $A$ in $\V$, $1^*A$ is an iteration semiring in $\W$.
\end{cor}

{\sl Proof.} Since the morphism $h$ is surjective, we have $1^*A \in \V$.
Since also $\b1^\otimes = 1^*1^{**} = 1^*1^* = 1^* = \b1$, 
it holds that $1^*A \in \W$. \eop

For the next corollary, note that if $t$ is a term over $\Sigma$ and 
$A$ is an iteration semiring, then $t$ induces a function 
$A^\Sigma \to A$ as usual. We will denote this function by $t^A$.
Below we will write function composition in the 
diagrammatic order.

\begin{cor}
Suppose that $t$ is a term over $\Sigma$ and $A \in \V$.
Then $(1^*t)^A$ can be factored
as $\overline{h} \circ (1^* t)^{1^* A}$, where $h$ is the morphism 
$A \to 1^*A$ of Lemma~\ref{lem-VW} and $\overline{h} : A^\Sigma \to (1^*A)^\Sigma$,
$e \mapsto e \circ h$.
\end{cor}

{\sl Proof.} Since $h$ is a morphism,  
$$\overline{h}\circ (1^*t)^{1^*A} = 
(1^* t)^A \circ h =
1^*(1^* t)^A =
(1^*1^*)t^A=
(1^*t)^A. \eop 
$$

\begin{cor}
\label{cor-equ}
Suppose that $s,t$ are terms over $\Sigma$ and $A \in \V$.
Then $1^*t = 1^*s$ holds in $A$ iff it holds in $1^*A$.
\end{cor}

We will use Theorem~\ref{thm-Krob} in the following way. 
Let $\Sigma$ be a set and consider a term $t$ over $\Sigma$.
It is easy to see by induction that if $1^*t$ evaluates to a series 
$r$ in $\B^{\rat} \llangle \Sigma^* \rrangle$, then in 
$\N_\infty^{\rat}\llangle \Sigma^* \rrangle$ it evaluates to 
the series whose nonzero coefficients are all $\infty$ and whose 
support is the same as that of $r$. 

We claim that if $1^*t$ and $1^* s$ evaluate 
to the same series in $\N_\infty^{\rat} \llangle \Sigma^* \rrangle$,
then the identity $1^* t = 1^* s$ holds in $\V$. 
Let $\W$ denote the variety of iteration semirings 
satisfying $1^* = 1$. Since $1^*t$ and $1^*s$ evaluate to the 
same series in $\N_\infty^{\rat} \llangle \Sigma^* \rrangle$,
they evaluate to the same series in $\B^{\rat} \llangle \Sigma^* \rrangle$.
Thus, by Theorem~\ref{thm-Krob}, $1^*t = 1^*s$ holds in $\W$. 
Let $A \in \V$. By Corollary~\ref{cor-VW}, $1^*A \in \W$, so 
$1^*t = 1^*s$ holds in $1^*A$. By Corollary~\ref{cor-equ}, this implies 
that $1^*t =1^*s$ holds in $A$. Since $A$ was an arbitrary 
iteration semiring in $\V$, this means that $1^*t = 1^* s$ holds in 
$\V$.

\begin{lem}
Suppose that $t,s$ are terms over $\Sigma$ such that the support of $|t|$
is included in the support of $|1^*s|$.
Then $t + 1^*s$ is equivalent to $1^* s$.
\end{lem}

{\sl Proof.}
By the above argument, $1^*s = 1^*(t+s) = 1^*t + 1^*s$ holds in $\V$.
Also, $t + 1^*t = (1 + 1^*)t = 1^*t$ holds. Thus,
$$t+ 1^*s = t + 1^*t +1^*s = 1^*t + 1^*s = 1^* s$$
holds. \eop 

We now prove a stronger version of Lemma~\ref{lem-3parts1}.

\begin{lem}
\label{lem-3parts2}
For every term $t$ there is an equivalent term of the form $t_c + t_0 + 1^*t_\infty$, 
where $t_c$ is a constant in $\N$, $t_0$ is an ideal term, and $t_\infty$ 
is a term. Moreover, if $t_c \neq 0$ then $t_\infty$ is ideal and $|t_0|$  
and $|1^*t_\infty|$ have disjoint supports.
\end{lem}

{\sl Proof.}
We know from Lemma~\ref{lem-3parts1} that $t$ is equivalent to a term of the form 
$t_c + t_0 + 1^*t_\infty$, where $t_c \in \N$, $t_0$ is an ideal term
and if $t_c$ is not $0$ then $t_\infty$ is also ideal. Now $\supp(1^*t_\infty) =
\supp(t_\infty)$ is a regular language which we denote by $R$. Consider the 
rational series $s_0 = |t_0|$ and write it as the sum $s_1 + s_2$, where 
$(s_1,w) = (s_0,w)$ if $w \not\in R$ and $(s_1,w) = 0$ otherwise, 
moreover, $(s_2,w) = (s_0,w)$ if $w \in R$ and $(s_2,w) = 0$ otherwise.
It is known that $s_1$ and $s_2$ are rational (see \cite{Berstel}) and thus 
there exsist ideal terms $t_1$ and $t_2$ with $|t_1| = s_1$ and $|t_2| = s_2$.
Since $|t_1 + t_2| = s_1 + s_2 = s = |t_0|$, and since these terms are ideal,
by Theorem~\ref{thm-free partial} we have that $t_0 = t_1 + t_2$ holds in $\V$.  
Since the support of $|t_2|$ is included in the 
support of $|1^*t_\infty|$,  $t_2 + 1^*t_\infty = 1^*t_\infty$ 
also holds in $\V$.
Summing up, $t$ is equivalent to $t_c + t_0 + 1^*t_\infty$ which is in turn
equivalent to $t_c + t_1 + 1^*t_\infty$ proving the claim. 
\eop

\begin{thm}
\label{thm-free1}
For each set $\Sigma$, $\N_\infty^{\rat}\llangle \Sigma^* \rrangle$ 
is freely generated by $\Sigma$ in $\V$.
\end{thm}

{\sl Proof.} 
We have already noted that $\N_\infty^{\rat}\llangle \Sigma^* \rrangle$ 
is in $\V$. By definition, $\Sigma$ generates 
$\N_\infty^{\rat}\llangle \Sigma^* \rrangle$.
But we still have to show that if two terms over $\Sigma$ 
evaluate to the same series, then they are equivalent. 
But any term $t$ is equivalent to some term 
of the form $t_c + t_0 + 1^*t_\infty$ where $t_c$ 
is a constant in $\N$ and $t_0$ is ideal, and if 
$t_c > 0$, then $t_\infty$ is ideal. 
Now $|t| = |t_c| + |t_0| + |1^*t_\infty|$, where 
$|t_c| \in \N$, $|t_0| \in \N\llangle \Sigma^* \rrangle$ 
and $|1^*t_\infty| \in \{0,\infty\}\llangle \Sigma^* \rrangle$,
i.e., each coefficient of the series $|1^* t_\infty|$ is 
$0$ or $\infty$. Moreover, $|t_0|$ and $|1^*t_\infty|$ have disjoint supports, 
and either $|t_c| = 0$ or $|1^*t_\infty|$ is proper. 
Thus, if $|t| = |s|$, then $|t_c| = |s_c|$, 
$|t_0| = |s_0|$ and  $|1^* t_\infty| = |1^* s_\infty|$.
By Corollary~\ref{cor-initial} we have that $t_c = s_c$ holds in $\V$.
Since $t_0$ and $s_0$ evaluate to the same series 
in $\N^{\rat}\llangle \Sigma^* \rrangle$, by Theorem~\ref{thm-free partial} 
we have that $t_0 = s_0$ holds in $\V$. Finally, by the above
discussion, $1^* t_\infty = 1^* s_\infty$ holds in $\V$,
proving that $t = s$ holds. \eop

\begin{cor}
\label{cor-split}
A series $s \in \N_\infty\llangle \Sigma^* \rrangle$ is in 
$\N_\infty^{\rat} \llangle \Sigma^* \rrangle$ iff $s = s_0 + s_\infty$ 
where $s_0 \in \N^{\rat}\llangle \Sigma^* \rrangle$ and all nonzero  
coefficients of $s_\infty \in \N_\infty^{\rat} \llangle \Sigma^* \rrangle$ are equal to $\infty$.
The series $s_0$ and $s_\infty$ may be chosen so that they have
disjoint supports. 
Moreover, a series $s$, all of whose nonzero coefficients are 
equal to $\infty$, is rational iff its support is regular.
\end{cor} 

\begin{remark}
The variety $\V$ is not finitely based, since it has a non-finitely based 
subvariety $\W$ which has a finite relative axiomatization over $\V$
by the single identity $1^* = 1$. See also \cite{Krob-Krat}. 
Likewise, the variety of all iteration semirings is non-finitely based. 
\end{remark}

Recall from \cite{Eilenberg} that a \emph{complete semiring} is a semiring $S$ which
is equipped with a summation operation $\sum_{i \in I} s_i$ for all 
index sets $I$ satisfying $\sum_{i \in \emptyset} = 0$, 
$\sum_{i \in \{1,2\}} s_i = s_ 1 + s_2$, moreover, product distributes 
over all sums and summation is associative: 
\begin{eqnarray*}
a(\sum_{i \in I} b_i) &=& \sum_{i \in I}ab_i\\
(\sum_{i \in I} b_i)a &=& \sum_{i \in I}b_ia\\
\sum_{j \in J}\sum_{i \in I_j}a_i &=& \sum_{i \in \cup_{j \in J}I_j}a_i,
\end{eqnarray*}
where in the last equation the sets $I_j$ are pairwise disjoint. 
\emph{Countably complete semirings} are defined in the same 
with the additional constraint that all sums are at most countable.
Clearly, every complete semiring is countably complete. 

An \emph{$\omega$-continuous} semiring \cite{BEbook} is a semiring $S$ equipped with a partial order
such that $S$ is an $\omega$-complete partial order ($\omega$-cpo) 
with bottom element $0$ and the sum and product operations are continuous, 
i.e., they preserve the suprema of $\omega$-chains. A \emph{continuous semiring}
is defined in the same way, it is a cpo with continuous operations such that 
$0$ is the bottom element. Each $\omega$-continuous semiring is a
countably complete semiring with 
\begin{eqnarray*}
\sum_{i \in I} s_i 
&=& 
 \sup\{\sum_{i\in F}s_i :
F \subseteq I\  {\rm finite} \}.
\end{eqnarray*} 
Similarly, each continuous semiring is 
complete. The semiring $\N_\infty$, equipped with the natural order, 
is continuous. It is well-known that equipped with the pointwise order, 
 $\N_\infty\llangle \Sigma^* \rrangle$ is also continuous for each $\Sigma$.  

When $S$ is countably complete, we can define a star operation on $S$ 
by $a^* = \sum_{n \geq 0} a^n$. Since $\omega$-continuous, continuous and 
complete semirings are all countably complete, the same definition 
applies to these semirings. We point out that the $^*$-semirings so obtained 
are all in $\V$. Indeed, it is known that when $S$ is countably complete,
then $S$ is an  iteration semiring (cf. \cite{BEbook}). 
We have that $1^*$ is a countable sum of $1$ with itself. 
Using distributivity, it follows that $1^*1^* = 1^*$. 
By distributivity, we also have (\ref{ax2}). Finally, $1^*a^*$ 
and $1^*(1^*a)^*$ are both equal to a countable sum $\sum_{i \in I} s_i$ 
containing for each $n$ a countable number of summands $s_i$ equal to $a^n$. 

By the above observations and the fact that the 
semirings $\N_\infty \llangle \Sigma^* \rrangle$ are continuous and 
contain the semirings $\N_\infty^{\rat} \llangle \Sigma^* \rrangle$,
we immediately have:

\begin{cor}
Continuous, $\omega$-continuous, complete and countably complete semi\-rings, 
equipped with the above star operation, satisfy exactly the identities 
of the variety $\V$.  
\end{cor}

\begin{remark}
The set $\N_\infty$ carries another important semiring structure.
Equipped with minimum as addition and addition as multiplication
(and $\infty$ as the additive identity element and $0$ as the 
multiplicative identity), $\N_\infty$ is called the 
\emph{tropical semiring}. It is known that the tropical
semiring has a non-finitely based equational theory,
cf. \cite{AEI}.  
 Krob \cite{Krob-tropical} has shown that 
the equality problem for  rational power series 
in two or more letters over the tropical semiring is undecidable.  
Rational power series in a single letter over the tropical semiring 
were treated in \cite{BonnierKrob}.
\end{remark}

We end this section by pointing out how Theorem~\ref{thm-Krob}
can be derived from Theorem~\ref{thm-free1}. When $k \geq 1$ is an integer,
let $\k$ denote the quotient of the iteration semiring $\N_\infty$
obtained by collapsing $k$ and $\infty$ and thus all elements 
of $\N_\infty$ at least $k$ . When $k = 1$, $\k$ is just the Boolean 
semiring $\B$ with star operation $0^* = 1^* = 1$. Our result is:

\begin{thm}
\label{thm-freek}
For each integer $k \geq 1$,\ $\k^{\rat}\llangle \Sigma^* \rrangle$ is freely 
generated by $\Sigma$ in the variety of iteration semiring satisfying 
the identity $1^* = k$.
\end{thm}

Of course, in the statement of the Theorem, $k$ also denotes the term 
$1+ \cdots + 1$ ($k$ times). Since any iteration semiring satisfying $1^* = k$ 
satisfies (\ref{ax1}), (\ref{ax2}) and (\ref{ax3}),
Theorem~\ref{thm-freek} is immediate from Theorem~\ref{thm-free1} if we can show that 
\begin{center}
the least congruence $\sim$ on $\N_\infty^{\rat}\llangle \Sigma^* \rrangle$ 
which collapses $k$ and $1^*$
\end{center}
 collapses any rational series in 
$\N_\infty^{\rat}\llangle \Sigma^* \rrangle$ with a series 
all of whose coefficients are either less than $k$  or equal to $1^* (= \infty)$. 
By Corollary~\ref{cor-split}, it is sufficient to prove 
this for rational series in 
$\N^{\rat}\llangle \Sigma^* \rrangle$.
The rest of this section is devoted to proving this fact.

\begin{lem}
Suppose that $s \in \N^{\rat} \llangle \Sigma^* \rrangle$ such that 
any nonzero coefficient of $s$ is at least $k$. Then $s \sim 1^*s$.
\end{lem}

{\sl Proof.} Let $R = \supp(s)$ which is a regular language in $\Sigma^*$
(cf. \cite{Berstel}),
and let $r$ denote the characteristic series of $R$, so that for any word
$w$, $(r,w) = 1$ if $w \in R$ and $(r,w) = 0$ otherwise. It is known 
that $r$ is rational (this is true for any semiring, cf. \cite{Berstel})
and thus $kr$ is also rational. Now it is known that $t = s -kr$ is also 
rational, see Theorem 1.8 in Chapter VII of \cite{Berstel}. 
It is clear that $kr \sim 1^*r = 1^*s$. Using this, we have:
$$s = kr + t \sim 1^*s + t = 1^* s. \eop$$

\begin{prop}
For each integer $k$ and each $s \in \N^{\rat}\llangle \Sigma^* \rrangle$ there is a series 
$r \in\N^{\rat}\llangle \Sigma^* \rrangle$ 
with $s \sim r$ such that all coefficients of $r$ are either less than $k$ 
or equal to $1^*$. 
\end{prop}

{\sl Proof.} In our argument, we will make use of the following known fact from
\cite{Berstel}. Given any rational series $s \in \N^{\rat}\llangle
\Sigma^* \rrangle$, $s$ can be written as a sum of 
rational series $s_0 + \cdots + s_k$ such that each coefficient of any 
$s_i$ with $i < k$ is $0$ or $i$, and each coefficient 
of $s_k$ is $0$ or $\geq k$. By the previous lemma, $s_k \sim 1^*s_k$,
and thus $s_k$ is congruent to the rational series $s_k'$ such that 
$(s_k',w) = 1^*$ if $(s_k,w) \geq k$ and $(s_k',w) = 0$ otherwise.  
We conclude that $s \sim s_0 + 
\cdots + s_{k-1} + s_k'$ which has the desired property. \eop

\section{A second characterization}
\label{sec-Ninftyrat2}

In the previous section, we have characterized the semi\-rings 
$\N_\infty^{\rat}\llangle \Sigma^* \rrangle$ as the free algebras 
in a non-finitely based variety $\V$ of $^*$-semi\-rings. 
Since $\N_\infty$ has a natural order, 
$\N_\infty \llangle \Sigma^* \rrangle$
may be equipped with the pointwise order. 
This order on $\N_\infty \llangle \Sigma^* \rrangle$
is actually the same as the sum order: For all
series $s,s' \in \N_\infty\llangle \Sigma^* \rrangle$, 
$s \leq s'$ iff there is a series $r$ with $s + r = s'$.
Moreover, since $\N_\infty$ is a continuous semi\-ring, 
cf. e.g. \cite{EsikKuichind}, so is $\N_\infty\llangle \Sigma^* \rrangle$. 
In particular, any map $x \mapsto sx + r$ over 
$\N_\infty\llangle \Sigma^* \rrangle$ has the series 
$s^*r$ as its least \emph{pre-fixed point} (since $ss^*r + r \leq s^*r$ 
and for all $s'$, if $ss' + r \leq s'$ then $s^*r \leq s'$).
Moreover, $rs^*$ is the least pre-fixed point of the map 
$x\mapsto xs + r$.
The semi\-ring $\N_\infty^{\rat} \llangle \Sigma^* \rrangle$,
equipped with the pointwise order inherited from 
$\N_\infty \llangle\Sigma^* \rrangle$
also has these least pre-fixed point properties. 
However, in the main result of this section, 
we will have to work with the sum order on $\N_\infty^{\rat}\llangle \Sigma^* \rrangle$
which is not the same as the pointwise order. It is known 
that for  $r,s \in \N^{\rat}\llangle \Sigma^*\rrangle$ 
with $r \leq s$ in the pointwise order, 
the difference $s - r$ may \emph{not} be rational (see \cite{Berstel}),
so that there may not exist a \emph{rational} series $r'$ with 
$r + r' = s$. Since $\N^{\rat}\llangle \Sigma^* \rrangle = 
\N_\infty^{\rat}\llangle \Sigma^* \rrangle \cap \N\llangle \Sigma^* \rrangle$,
the same holds for $\N_\infty^{\rat}\llangle \Sigma^* \rrangle$. 
But the above least pre-fixed point property still holds 
in $\N_\infty^{\rat}\llangle \Sigma^* \rrangle$ with the sum order,
as will be shown below.

For the rest of this paper, by an \emph{ordered semi\-ring}
we shall mean a semi\-ring $S$ equipped with a partial order 
$\leq$ preserved by  sum and product: 
If $a \leq a'$ and $b \leq b'$ then $a + b \leq a' + b'$ and 
$ab \leq a'b'$. Following \cite{EsikKuichind}, we call a $^*$-semi\-ring an 
\emph{inductive $^*$-semi\-ring} if it is an ordered semi\-ring 
such that the following hold for all $a,b,x \in S$:
\begin{eqnarray}
&&aa^* + 1 \leq a^*\\
&&ax + b \leq x \quad \Rightarrow \quad a^*b \leq x.
\end{eqnarray} 
It then follows that for any $a,b$, $a^*b$ is the least pre-fixed point 
of the map $x \mapsto ax + b$, and is actually a fixed point.
Moreover, it is known that the star operation is 
also monotone in any inductive $^*$-semi\-ring. 
A \emph{symmetric inductive $^*$-semi\-ring} $S$ also 
satisfies 
\begin{eqnarray}
xa + b \leq x &\Rightarrow & ba^* \leq x
\end{eqnarray}  
for all $a,b,x \in S$. In \cite{Kozen}, Kozen defines a \emph{Kleene algebra}
as an idempotent symmetric inductive $^*$-semi\-ring. (Note that  
if an ordered semiring $S$ is idempotent, then the partial order 
is the semilattice order: $a \leq b$ iff $a + b = b$.)
A morphism of (symmetric) inductive $^*$-semi\-rings is 
a $^*$-semi\-ring morphism which preserves the order.

The following result was proved in \cite{EsikKuichind}:

\begin{thm}
\label{thm-ind-it}
Every inductive $^*$-semiring is an iteration semiring 
satisfying $1^* = 1^{**}$.
\end{thm}

We call a (symmetric)  inductive 
$^*$-semi\-ring  \emph{sum ordered}, if its order relation is given by
$a \leq b$ iff there is some $c$ with $a+c = b$.
Suppose that $S$ is an inductive $^*$-semi\-ring. Since for any $x \in S$,
$1x +0 = x$, we have that $0 = 1^*0 \leq x$. Thus, $0$ is the least 
element of $S$ and since the order is preserved by addition, 
$x \leq x+y$ for all $x,y \in S$, so that the order on $S$ is 
an extension of the sum order. 

\begin{prop}
Any inductive $^*$-semi\-ring satisfies (\ref{ax1}) and 
(\ref{ax3}), i.e., the identities $1^*1^* = 1^*$ and $1^*(1^*a)^* = 1^*a^*$.
Moreover, any inductive $^*$-semi\-ring satisfies $1^*a \leq a1^*$.
\end{prop}

{\sl Proof.} The identity $1^* = 1^*1^*$ holds by Theorem~\ref{thm-ind-it} 
and the description of the initial iteration semiring.

Now for the inequality $1^* a \leq a1^*$. We have 
$1(a1^*) + a = a(1^* + 1) = a1^*$. Thus, $1^* a \leq a1^*$.

Last  for (\ref{ax3}). 
On one hand, $a \leq 1^*a$, and thus  $a^* \leq (1^*a)^*$ and 
$1^* a^* \leq 1^* (1^* a)^*$.
On the other hand, $1(1^*(1^*a)^*) + a^* \leq 1^*(1^*a)^* + (1^* a)^*
= (1^* + 1)(1^*a)^* = 1^* (1^* a)^*$, and thus $1^* a^* 
\leq 1^*(1^* a)^*$. \eop

\begin{prop}
In any symmetric inductive $^*$-semi\-ring $S$, 
$1^* a =a 1^*$ for all $a \in S$.
\end{prop}

{\sl Proof.} We have seen that $1^*a \leq a1^*$. Since 
$(1^*a)1 + a = (1^* + 1)a = 1^* a$, it holds that $a1^* \leq 1^*a$. \eop

Consider now $\N_\infty^{\rat}\llangle \Sigma^* \rrangle$ equipped 
with the \emph{sum order}, denoted $\leq$. We claim that  
$\N_\infty^{\rat}\llangle \Sigma^* \rrangle$ is a symmetric
inductive $^*$-semiring. In \cite{EsikKuichind}, it is shown that 
if an ordered semiring equipped with 
a star operation is ordered by the sum order,
then it is a symmetric inductive $^*$-semiring iff 
it satisfies $aa^* + 1 = a^*$ and 
\begin{eqnarray*}
&&ax + b = x \quad \Rightarrow \quad a^*b \leq x\\
&&xa + b = x \quad \Rightarrow \quad ba^* \leq x.
\end{eqnarray*}
In order to prove these properties hold, we describe all solutions 
of a linear fixed point equation over 
$\N_\infty\llangle \Sigma^* \rrangle$.

\begin{prop}
Let $s,r$ be  series in $\N_\infty\llangle \Sigma^* \rrangle$.
\begin{enumerate}
\item
If $s$ is proper, then the equation $x =sx + r$ has $s^*r$ as its unique solution.
\item 
If $s = 1 + s_0$, where $s_0$ is proper, then the solutions of $x =sx + r$ 
are the series of the form $s^*r + 1^*s_0^+ t + t$, where $t$ is any series.
\item 
If $s = k + s_0$, where $s_0$ is proper and $k \in \N_\infty$, $k \neq 0,1$, then the 
solutions of $x = sx  + r$ are the series of the form $s^*r + 1^*s_0^*t =
s^*(r+t)$, where $t$ is any series.
\end{enumerate}
\end{prop}

{\sl Proof.} For the first claim, see \cite{Berstel}. Assume that $s = k + s_0$ 
where $s_0$ is proper and $k \in \N_\infty$, $k \neq 0$. Consider the equation
$x = f(x)$, where $f(x) = sx + r$. Since $\N_\infty$ 
is continuous and $x\leq f(x)$, all solutions can be 
obtained by starting with a series $t$ and forming the increasing 
sequence $f^n(t)$, for $n \geq 0$, and taking the supremum of this
sequence. Since $f^n(t) = s^nt + s^{n-1}r + \cdots + r$, this gives
$s^*r + \sup_{n \geq 0} s^n t$. But for each $n$, $s^nt = (k + s_0)^nt$,
and using the expansion $(k + s_0)^n = k^n   + 
\left(\begin{array}{c} n\\ 1 \end{array}\right) k^{n-1} s_0 + 
\left(\begin{array}{c} n\\ 2 \end{array}\right) k^{n-2} s_0^2 + 
\cdots + s_0^n$, we obtain that
\begin{eqnarray*}
\sup_{n \geq 0} s^n t
&=& 
\left\{
\begin{array}{ll}
1^*s_0^+t + t & {\rm if}\ k = 1\\
1^*s_0^*t & {\rm if}\ k > 1.
\end{array}
\right.
\end{eqnarray*}
\eop

\begin{cor}
Let $s,r$ be  series in $\N_\infty^{\rat}\llangle \Sigma^* \rrangle$
and consider the equation $x = sx + r$ with least solution $s^*r$.
If $z$ is any rational solution, then there is a rational series 
$p \in \N_\infty^{\rat}\llangle \Sigma^* \rrangle$ with $z = s^*r + p$.
\end{cor} 

{\sl Proof.} When $s$ is proper, $s^*r$ is the only solution and our claim is clear.
Assume that $s = 1 + s_0$, where $s_0$ is proper (and rational). Then the least solution 
is $s^*r = 1^*s_0^*r$, and any other solution
$z$ is of the form $z = s^*r + 1^*s_0^+t + t = 1^*s_0^*r + 1^*s_0^+t + t$.
We see that $z = s^*r + z$, so if $z$ is rational, then it is the sum of the 
least solution with a rational series. The last case is when $s = k + s_0$,  
where $s_0$ is proper and $k \in \N_\infty$, $k > 1$. Then the least 
solution is $s^*r = 1^*s_0^*r$, and any other solution
$z$ is of the form $z = 1^* s_0^* r + 1^*s_0^*t$. We again have $z = s^*r + z$,  
so that if $z$ is rational, then it is the sum of the least solution 
with a rational series. \eop 

Symmetrically, we have: 

  \begin{cor}
Let $s,r$ be  series in $\N_\infty^{\rat}\llangle \Sigma^* \rrangle$
and consider the equation $x = xs + r$ with least solution $rs^*$.
If $z$ is any rational solution, then there is a rational series 
$p \in \N_\infty^{\rat}\llangle \Sigma^* \rrangle$ with $z = rs^* + p$.
\end{cor}  

The main result of this section is:

\begin{thm}
\label{thm-free2}
For each $\Sigma$, $\N_\infty^{\rat}\llangle \Sigma^* \rrangle$ 
(equipped with the sum order) is 
freely generated by $\Sigma$ in the class of all inductive $^*$-semi\-rings
satisfying $a1^* \leq 1^*a$. In detail, 
for any inductive $^*$-semi\-ring $S$ satisfying $a1^* \leq 1^*a$ and 
for any function $h: \Sigma \to S$ there is a unique inductive $^*$-semi\-ring 
morphism $h^\sharp: \N_\infty\llangle \Sigma^* \rrangle \to S$ extending $h$.
\end{thm}

{\sl Proof.} 
We have already proved that $\N_\infty^{\rat}\llangle \Sigma^* \rrangle$
equipped with the sum order is an inductive $^*$-semi\-ring
satisfying $a1^* \leq 1^*a$.
Suppose that $S$ and $h$ are given. Since $S$ 
satisfies $a1^* \leq 1^*a$, $S$ is an iteration semi\-ring in the variety
$\V$ described in the previous section. By Theorem~\ref{thm-free1}, $h$ extends 
to a morphism $h^\sharp$ of $^*$-semi\-rings. To see that $h^\sharp$ 
is preserves the order, assume that $s,s' \in \N_\infty^{\rat}\llangle
\Sigma^*\rrangle$ with $s \leq s'$. Then there exists a \emph{rational}
series $r$ in  $ \N_\infty^{\rat}\llangle \Sigma^*\rrangle$ with
$s + r = s'$. Since $h^\sharp$ preserves $+$, also $sh + rh = s'h$.
But the order on $S$ contains the sum order, so that $sh \leq s'h$. 
The fact that $h^\sharp$ is unique follows from Theorem~\ref{thm-free1}.
\eop   

\begin{cor}
For each $\Sigma$, $\N_\infty^{\rat}\llangle \Sigma^* \rrangle$ is 
freely generated by $\Sigma$ in the class of all 
symmetric inductive $^*$-semi\-rings.
\end{cor}

\begin{cor}
For each $\Sigma$, $\N_\infty^{\rat}\llangle \Sigma^* \rrangle$ is 
freely generated by $\Sigma$ in the class of all sum ordered 
(symmetric) inductive $^*$-semi\-rings.
\end{cor}



Let $k \geq1$ and consider the $^*$-semiring $\k$. Equipped with the natural order,
$\k$ is a continuous semiring and thus $\k\llangle \Sigma^* \rrangle$ is also 
a continuous semiring and a symmetric inductive $^*$-semiring. We can write each 
series $s \in \k\llangle \Sigma^* \rrangle$ in a unique way as the sum of 
series $s_0,s_1,\cdots,s_k$, where each nonzero coefficient in any $s_i$ is $i$. 
(Of course, $s_0$ is the $0$ series.) Now by finiteness, it is known that 
$s$ is rational iff each $s_i$ is rational iff the support of 
each $s_i$ is regular. Using this, we do not have the problem encountered 
in connection with the ordering of $\N_\infty^{\rat}\llangle \Sigma^* \rrangle$, 
the sum order and the pointwise order are equivalent on 
$\k^{\rat}\llangle \Sigma^* \rrangle$. Using Theorem~\ref{thm-freek} 
we have:

\begin{thm}
For each $k \geq 1$, $\k^{\rat}\llangle \Sigma^* \rrangle$ is both the free 
inductive $^*$-semiring and the free symmetric inductive $^*$-semiring 
on the set $\Sigma$ satisfying the identity $k = k+1$.
\end{thm}

{\sl Proof.} Let $S$ be an inductive $^*$-semiring satisfying $k = k+1$.
Since $1 \leq 1^*$ in $S$, also $k \leq 1^*$. Since $k + 1 = k$, 
also $1^* \leq k$. Thus $1^* = k$. By Theorem~\ref{thm-freek}, every 
function $\Sigma \to S$  extends to a morphism 
$h^\sharp : \k^{\rat}\llangle \Sigma^* \rrangle \to S$ of iteration semirings.
The extension $h^\sharp$ is monotone (and unique). \eop

\end{document}

%% file: mymacs.tex
\def\adjusttextwidth#1{
	\newdimen\hdiff
	\hdiff = #1
	\advance\textwidth by \hdiff
	\divide\hdiff by -2
	\advance\hoffset by \hdiff
}

\def\adjusttextheight#1{
	\newdimen\vdiff
	\vdiff = #1
	\advance\textheight by \vdiff
	\divide\vdiff by -2
	\advance\voffset by \vdiff
}

\catcode`\@=11  
\def\newrmtheorem#1{\@ifnextchar[{\@ormthm{#1}}{\@nrmthm{#1}}}

\def\@nrmthm#1#2{%
\@ifnextchar[{\@xnrmthm{#1}{#2}}{\@ynrmthm{#1}{#2}}}

\def\@xnrmthm#1#2[#3]{\expandafter\@ifdefinable\csname #1\endcsname
{\@definecounter{#1}\@addtoreset{#1}{#3}%
\expandafter\xdef\csname the#1\endcsname{\expandafter\noexpand
  \csname the#3\endcsname \@thmcountersep \@thmcounter{#1}}%
\global\@namedef{#1}{\@rmthm{#1}{#2}}\global\@namedef{end#1}{\@endtheorem}}}

\def\@ynrmthm#1#2{\expandafter\@ifdefinable\csname #1\endcsname
{\@definecounter{#1}%
\expandafter\xdef\csname the#1\endcsname{\@thmcounter{#1}}%
\global\@namedef{#1}{\@rmthm{#1}{#2}}\global\@namedef{end#1}{\@endtheorem}}}

\def\@ormthm#1[#2]#3{\expandafter\@ifdefinable\csname #1\endcsname
  {\global\@namedef{the#1}{\@nameuse{the#2}}%
\global\@namedef{#1}{\@rmthm{#2}{#3}}%
\global\@namedef{end#1}{\@endtheorem}}}

\def\@rmthm#1#2{\refstepcounter
    {#1}\@ifnextchar[{\@yrmthm{#1}{#2}}{\@xrmthm{#1}{#2}}}

\def\@xrmthm#1#2{\@beginrmtheorem{#2}{\csname the#1\endcsname}\ignorespaces}
\def\@yrmthm#1#2[#3]{\@opargbeginrmtheorem{#2}{\csname
       the#1\endcsname}{#3}\ignorespaces}

\def\@beginrmtheorem#1#2{\trivlist \item[\hskip \labelsep{\bf #1\ #2}]}
\def\@opargbeginrmtheorem#1#2#3{\trivlist
      \item[\hskip \labelsep{\bf #1\ #2\ (#3)}]}

\catcode`\@=12  

\catcode`\@=11  
\def\newsmtheorem#1{\@ifnextchar[{\@osmthm{#1}}{\@nsmthm{#1}}}

\def\@nsmthm#1#2{%
\@ifnextchar[{\@xnsmthm{#1}{#2}}{\@ynsmthm{#1}{#2}}}

\def\@xnsmthm#1#2[#3]{\expandafter\@ifdefinable\csname #1\endcsname
{\@definecounter{#1}\@addtoreset{#1}{#3}%
\expandafter\xdef\csname the#1\endcsname{\expandafter\noexpand
  \csname the#3\endcsname \@thmcountersep \@thmcounter{#1}}%
\global\@namedef{#1}{\@smthm{#1}{#2}}\global\@namedef{end#1}{\@endtheorem}}}

\def\@ynsmthm#1#2{\expandafter\@ifdefinable\csname #1\endcsname
{\@definecounter{#1}%
\expandafter\xdef\csname the#1\endcsname{\@thmcounter{#1}}%
\global\@namedef{#1}{\@smthm{#1}{#2}}\global\@namedef{end#1}{\@endtheorem}}}

 \def\@osmthm#1[#2]#3{\expandafter\@ifdefinable\csname #1\endcsname
  {\global\@namedef{the#1}{\@nameuse{the#2}}%
\global\@namedef{#1}{\@smthm{#2}{#3}}%
\global\@namedef{end#1}{\@endtheorem}}}

\def\@smthm#1#2{\refstepcounter
    {#1}\@ifnextchar[{\@ysmthm{#1}{#2}}{\@xsmthm{#1}{#2}}}

\def\@xsmthm#1#2{\@beginsmtheorem{#2}{\csname the#1\endcsname}\ignorespaces}
\def\@ysmthm#1#2[#3]{\@opargbeginsmtheorem{#2}{\csname
       the#1\endcsname}{#3}\ignorespaces}

\def\@beginsmtheorem#1#2{\footnotesize \trivlist \item[\hskip \labelsep{\bf #1\ #2}]}
\def\@opargbeginsmtheorem#1#2#3{\footnotesize \trivlist
      \item[\hskip \labelsep{\bf #1\ #2\ (#3)}]}

\catcode`\@=12  

\newtheorem{thm}{\sc Theorem}[section]
\newtheorem{lem}[thm]{\sc Lemma}        
\newtheorem{prop}[thm]{\sc Proposition}  
\newtheorem{cor}[thm]{\sc Corollary}
\newtheorem{deff}[thm]{\sc Definition}

\newsmtheorem{exer}{\sc Exercise}[section]  
\newsmtheorem{expl}[thm]{\sc Example}       
\newrmtheorem{remark}[thm]{\sc Remark}
\newrmtheorem{problem}{\sc Problem}[section]

\catcode`\@=11
\font\teneuf=eufm10
\font\seveneuf=eufm7
\font\fiveeuf=eufm5
\newfam\euffam
\textfont\euffam=\teneuf
\scriptfont\euffam=\seveneuf
\scriptscriptfont\euffam=\fiveeuf
\def\frak{\let\next\relax\ifmmode\let\next\frak@\else
\def\next{\errormessage{Use \string\frak\space only in math mode}}\fi\next}
\def\goth{\let\next\relax\ifmmode\let\next\frak@\else
\def\next{\errormessage{Use \string\goth\space only in math mode}}\fi\next}
\def\frak@#1{{\frak@@{#1}}}
\def\frak@@#1{\noaccents@\fam\euffam#1}
\def\noaccents@{\def\accentfam@{0}}
\def\accentfam@{7}
\catcode`\@=12

\def\hexnumber@#1{\ifcase#1 0\or1\or2\or3\or4\or5\or6\or7\or8\or9\or
A\or B\or C\or D\or E\or F\fi}

\def\ST{
\ifmmode \Sigma {\rm Term}
\else  $ \Sigma {\rm Term} $ \fi }

\def\tuple#1#2#3{
\ifmmode #1_{#2}, \ldots, #1_{#3} \!
\else  $ #1_{#2}, \ldots, #1_{#3} \! $ \fi }

\def\atuple#1#2#3{
\hbox{$ \langle \kern -.1em #1_{#2}, \ldots, #1_{#3} \kern -.1em \rangle $} }

\def\ptuple#1#2#3{
\ifmmode ( #1_{#2}, \ldots, #1_{#3} ) 
\else  $ ( #1_{#2}, \ldots, #1_{#3} ) $  \fi }

\def\btuple#1#2#3{
\ifmmode \enskip [ #1_{#2}, \ldots, #1_{#3} ] \enskip
\else  $ [ #1_{#2}, \ldots, #1_{#3} ] $  \fi }

\def\range#1#2#3{
\ifmmode \enskip #1 = #2, \ldots, #3 \enskip
\else  $ #1 = #2, \ldots, #3 $ \fi }

\def\Pow#1{
   \hbox{\kern -.05em \bf Pow$_{#1} \kern -.06em$}}

\def\Pfn#1{
 \hbox{\kern -.05em \bf Pfn$_{#1}\kern -.06em$}}

\def\Rel#1{
 \hbox{\kern -.05em \bf Rel$_{#1}\kern -.06em$}}

\def\SeqRel#1{
 \hbox{\kern -.05em \bf SeqRel$_{#1}\kern -.06em$}}

 \def\Seq#1{
 \hbox{\bf Seq$_{#1}\kern -.06em$}}

\def\1#1{
\hbox{{\bf 1}$_{#1}$}
}

\def\dot{
\hbox{$ \kern -.1em  \cdot \kern -.2em $} }

\def\arr#1#2#3{
\hbox{$ \kern -.2em #1 \, \colon #2 \kern -.1em \rightarrow #3 \kern -.1em $}
}
\def\arrp#1#2#3{
\hbox{$ \kern -.2em #1 \, \colon #2 \kern -.1em \rightarrow #3 \kern -.1em .$}
}

\def\PS#1#2{
\hbox{$ #1 \langle \kern -.25pt  \langle #2 ^* \rangle \kern -.25pt \rangle$} }
 
\def\rat{
\hbox{\small rat} }

\def\ccirc{
\mathrel { \raise .5pt \hbox{$\scriptstyle \circ $}  }}

\def\Bbox{
{\unskip\nobreak\hfil\penalty50
\hskip1em\hbox{}\nobreak\hfil{\lower .5pt \hbox{$\Box$}}
\parfillskip=0pt \finalhyphendemerits=0 \par}
}

\def\eop{
\ifmmode {\hbox{\Bbox}} \else \Bbox \fi
}

\def\bbox{
\ifmmode {\hbox{\bbox}} \else \Bbox \fi
}

\def\ol#1{\overline{#1}}

\def\ii{
\hbox{$I \kern -.36em I$}}

\def\zz{\hbox{$Z \kern -.54em Z$}}

\def\nn{\hbox{$I \kern -.38em N$}}

\def\qq{\hbox{$I \kern -.55em Q$}}

\def\B{\hbox{${\cal B}$}}

\def\N{\rm \bf N}

\def\V{\hbox{${\cal V}$}}
